%% file: submission.tex
\documentclass[11pt]{article}

\usepackage[a4paper,margin=1in]{geometry}
\usepackage[T1]{fontenc}
\usepackage{lmodern}
\usepackage{microtype}
\usepackage{amsthm}
\usepackage{aliascnt}

\newcommand{\paperroot}{}
\IfFileExists{arxiv_ready/macros.tex}{\renewcommand{\paperroot}{arxiv_ready/}}{}
\input{\paperroot macros}

\usepackage[hidelinks]{hyperref}
\hypersetup{
  pdftitle={Parameterized Approximation for Minimizing Uncovered Elements and Unsatisfied Constraints},
  pdfauthor={Rudranarayan Kar, Souvik Saha, Abhishek Sahu},
  pdfsubject={Parameterized approximation of coverage and Boolean constraint satisfaction},
  pdfkeywords={parameterized approximation, minimum uncoverage, constraint satisfaction, semi-ladder index}
}

\newtheorem{theorem}{Theorem}[section]
\newaliascnt{lemma}{theorem}
\newtheorem{lemma}[lemma]{Lemma}
\aliascntresetthe{lemma}
\newaliascnt{corollary}{theorem}
\newtheorem{corollary}[corollary]{Corollary}
\aliascntresetthe{corollary}
\newaliascnt{proposition}{theorem}
\newtheorem{proposition}[proposition]{Proposition}
\aliascntresetthe{proposition}
\newaliascnt{observation}{theorem}
\newtheorem{observation}[observation]{Observation}
\aliascntresetthe{observation}
\theoremstyle{definition}
\newaliascnt{definition}{theorem}
\newtheorem{definition}[definition]{Definition}
\aliascntresetthe{definition}
\theoremstyle{remark}

\usepackage[capitalise,noabbrev]{cleveref}
\crefname{lemma}{Lemma}{Lemmas}
\crefname{corollary}{Corollary}{Corollaries}
\crefname{proposition}{Proposition}{Propositions}
\crefname{observation}{Observation}{Observations}
\crefname{definition}{Definition}{Definitions}
\crefname{claim}{Claim}{Claims}
\newaliascnt{claim}{theorem}
\newtheorem{claim}[claim]{Claim}
\aliascntresetthe{claim}
\title{\vspace{-2em}Parameterized Approximation for Minimizing Uncovered Elements and
Unsatisfied Constraints}
\author{Rudranarayan Kar\textsuperscript{1}\quad
Souvik Saha\textsuperscript{2}\quad
Abhishek Sahu\textsuperscript{1}\\[0.5em]
\small\textsuperscript{1}School of Computer Sciences, National Institute of Science Education\\
\small and Research (NISER), Bhubaneswar, India\\
\small (A constituent institution of Homi Bhabha National Institute (HBNI), Mumbai)\\
\small\textsuperscript{2}Chennai Mathematical Institute, Chennai, India}
\date{}

\begin{document}
\raggedbottom
\maketitle

\begin{abstract}
\input{\paperroot sections/abstractnew}

\end{abstract}

\input{\paperroot sections/introv2}
\input{\paperroot sections/preliminaries}
\input{\paperroot sections/bounded_arity_csp}
\input{\paperroot sections/kdd_hardness}
\input{\paperroot sections/semiladder_uncoverage}
\input{\paperroot sections/cardinality_augmentation}
\input{\paperroot sections/conclusion}

\bibliographystyle{plainurl}
\bibliography{\paperroot references}

\appendix
\input{\paperroot sections/min_uncoverage}
\input{\paperroot sections/bounded_set_size}

\end{document}

%% file: macros.tex
\usepackage{amsmath,amssymb,mathtools}
\usepackage{algorithm}
\usepackage[noend]{algpseudocode}
\usepackage{booktabs}
\usepackage{float}
\usepackage{tabularx}
\newcommand{\FPT}{\textsf{FPT}}
\newcommand{\Wone}{\textsf{W[1]}}
\newcommand{\Wtwo}{\textsf{W[2]}}
\newcommand{\MinUnCoverage}{\textnormal{\textsc{Min-UnCoverage}}}
\newcommand{\MinUnsatSAT}{\textnormal{\textsc{Min-Unsat SAT}}}
\newcommand{\MinUnsatCSP}{\textnormal{\textsc{Min-Unsat CSP}}}
\newcommand{\OPT}{\operatorname{OPT}}
\newcommand{\unsat}{\operatorname{unsat}}
\newcommand{\miss}{\operatorname{miss}}
\newcommand{\cov}{\operatorname{cov}}
\newcommand{\occ}{\operatorname{occ}}
\newcommand{\zero}{\mathbf{0}}
\newcommand{\eps}{\varepsilon}
\algnewcommand{\Input}{\item[\algorithmicrequire]}
\algnewcommand{\Output}{\item[\algorithmicensure]}

\makeatletter
\AtBeginDocument{%
  \providecommand{\theHALG@line}{\thealgorithm.\arabic{ALG@line}}}
\makeatother

%% file: sections/abstractnew.tex
We study three weighted residual-minimization problems under a cardinality
bound $k$. In \textsc{Min-Unsat CSP}, one sets at most $k$ Boolean variables
to true and minimizes the weight of the unsatisfied constraints. Restricting
the constraints to clauses gives \textsc{Min-Unsat SAT}. Its monotone special
case is equivalent to \textsc{Min-UnCoverage}, in which one selects at most
$k$ sets and minimizes the weight of the uncovered elements. Although the complementary
maximization problems have the same optimal solutions, their approximation
guarantees do not transfer to the residual objectives: when $k$ sets cover every
element, a constant-factor approximation for \textsc{Maximum Coverage} may
still leave elements uncovered, whereas every finite-factor approximation for
\textsc{Min-UnCoverage} must find a complete cover.  Indeed,
\textsc{Maximum Coverage} admits a polynomial-time
$(1-1/e)$-approximation, but by the standard $\mathsf{W[2]}$-hardness of
\textsc{Set Cover}, unless $\mathsf{FPT}=\mathsf{W[2]}$, none of the three unrestricted minimization
problems admits a finite-factor FPT approximation parameterized by $k$.

We identify structural conditions under which this obstruction can be
overcome.
For constraints of arity at most $d$, we give a deterministic
$(1+\varepsilon)$-approximation for weighted \textsc{Min-Unsat CSP} running
in $(dk/\varepsilon)^{O(k)}N^{O(1)}$
time, where $N$ is the input size. This also applies to bounded-arity
\textsc{Min-Unsat SAT} and to \textsc{Min-UnCoverage} when every element
belongs to at most $d$ sets. Bounding variable frequency, which corresponds
to set size for coverage, yields a different picture. Weighted
\textsc{Min-UnCoverage} is exactly solvable in
$2^{O(kd)}N^{O(1)}$ time when every set has size at most $d$. In contrast, we show that
unless $\mathsf{FPT}=\mathsf{W[1]}$, \textsc{Min-Unsat SAT} and
\textsc{Min-Unsat CSP} admit no finite-factor FPT approximation when every
variable occurs in at most $d$ clauses or constraints. Under the Exponential
Time Hypothesis (ETH), no such
approximation runs in $f(k,d)N^{o(\sqrt{k+d})}$ time for any computable $f$.

We next consider incidence-graph restrictions that generalize these two
one-sided degree bounds. The incidence graph joins selectable objects, sets or
variables, to elements, clauses, or constraints through membership or
occurrence. We give a deterministic efficient parameterized approximation
scheme (EPAS) for weighted
\textsc{Min-UnCoverage} on incidence graphs of semi-ladder index at most $d$,
running in time $(dk/\varepsilon)^{
    O((d^2k/\varepsilon)\log(2k/\varepsilon))
  }N^{O(1)}$.
 In contrast, unless
$\mathsf{FPT}=\mathsf{W[1]}$, unweighted \textsc{Min-Unsat SAT} and
\textsc{Min-Unsat CSP} admit no finite-factor FPT approximation even on
$K_{3,3}$-free incidence graphs and incidence graphs with semi-ladder index at most four. Under ETH,
no such approximation runs in $f(k,d)N^{o(k+d)}$ time, where $d$ denotes the
corresponding structural bound.

Bounded semi-ladder index implies bounded VC dimension. Bounded VC dimension,
however, marks where our positive approximation results stop. Unlike
\textsc{Maximum Coverage}, which admits an EPAS on this class, none of our
three residual-minimization problems admits a finite-factor FPT approximation
even on unweighted instances of VC dimension two, unless
$\mathsf{FPT}=\mathsf{W[1]}$.

Finally, we ask whether the cardinality bound can be relaxed while preserving
an upper bound on the uncovered or unsatisfied weight. For coverage on
incidence graphs of semi-ladder index at most $d$, a known FPT algorithm
preserves the target using at most $k+1$ sets whenever $k$ sets suffice. In
contrast, we prove that, unless $\mathsf{FPT}=\mathsf{W[1]}$, no FPT algorithm
for unweighted \textsc{Min-Unsat SAT} or \textsc{Min-Unsat CSP} can replace
$k$ by any computable bound $g(k,d)\geq k$ while preserving the target. This holds
even when every clause or constraint has arity at most three and, separately,
when every variable has frequency at most $d$.

%% file: sections/introv2.tex
\section{Introduction}

\label{sec:introduction}

Coverage and satisfaction problems under a cardinality bound can be viewed
from two complementary perspectives:
maximizing the weight of the elements, clauses, or constraints that are
covered or satisfied, or minimizing the weight of those that remain uncovered
or unsatisfied.  Because these two objective values sum to the total input
weight, the formulations have the same optimal solutions.  Their multiplicative approximation guarantees, however,
do not translate into one another.  When an optimal solution leaves only a
small amount of weight uncovered or unsatisfied, a constant-factor
approximation to the covered or satisfied weight may leave a residual
arbitrarily larger than the optimum, whereas an approximation to the residual
controls this quantity directly.  Conversely, when the optimal covered or
satisfied weight is small, an approximation to the residual may provide
little information about the weight actually covered or satisfied, making
the maximization viewpoint more informative.  Thus, depending on the
instance, either side of the complementary objective may provide the more
meaningful approximation guarantee.  While the maximization viewpoint has
been studied extensively for coverage and satisfiability problems, these
differences motivate a systematic study of the corresponding
residual-minimization objectives under a cardinality bound.

The classical \textsc{Maximum Coverage} problem asks for at most $k$ sets
whose union has maximum total weight.  We study the complementary residual
objective: selecting at most $k$ sets so as to minimize the total weight of
the elements left uncovered.  This objective arises naturally when the
unserved portion is itself the quantity of interest.  For example, under the
approval-based Chamberlin--Courant rule, each candidate corresponds to the
set of voters who approve that candidate.  Thus, after selecting a committee,
the uncovered elements correspond precisely to voters who approve none of
its members~\cite{chamberlin1983representative,skowron2017boundedfrequencies}. We call this problem \MinUnCoverage{}.  It has previously appeared under the
names \textsc{MinNonCovered}~\cite[Definition~2.2]{skowron2017boundedfrequencies}
and \textsc{Min $k$-Uncovered}~\cite[Section~4.2.3]{feldmann2020parameterizedsurvey}.
Formally, an instance consists of a ground set $U$, a family
$\mathcal S\subseteq 2^U$, positive rational weights on the ground elements,
and a cardinality bound $k$.  The goal is to select at most $k$ sets while minimizing
the total weight of the elements they leave uncovered.
We study this objective together with two satisfaction
analogues.
An instance of \MinUnsatSAT{} contains a CNF formula with positive
rational clause weights, and an instance of \MinUnsatCSP{} contains weighted
constraints over Boolean variables.  In both problems, the goal is to set at
most $k$ variables to true while minimizing the total weight of the
unsatisfied clauses or constraints.  All weights are encoded in binary.  The
three problems form a hierarchy that preserves objective weights:
\MinUnCoverage{} is the monotone special case of \MinUnsatSAT{}, which is a
special case of \MinUnsatCSP{}.
The cardinality bound is an upper bound throughout; the distinction from
requiring exactly $k$ selected objects is discussed in
\cref{prop:exact-cardinality}.

For coverage, the separation between the two approximation
viewpoints \textsc{Maximum Coverage} and \MinUnCoverage{} is already extreme.  As Skowron and Faliszewski emphasize, when $k$ sets can
cover the whole universe, a constant-factor approximation for \textsc{Maximum Coverage}
may still miss a constant fraction of the elements, whereas every finite-factor
approximation for \MinUnCoverage{} must find a complete cover
\cite[p.~692]{skowron2017boundedfrequencies}.  On the maximization side, the
classical greedy algorithm achieves a $(1-1/e)$ approximation
\cite{nemhauser1978submodular}, and Feige proved that, for every constant
$\delta>0$, no polynomial-time algorithm can achieve $1-1/e+\delta$ unless
$\mathsf{P}=\mathsf{NP}$~\cite{feige1998threshold}.
Even allowing
parameterized running time does not improve this threshold in general: under
the randomized Gap Exponential Time Hypothesis (Gap-ETH), no
$T(k)N^{o(k)}$-time algorithm achieves
$1-1/e+\delta$ for any function $T$ and any constant $\delta>0$
\cite{manurangsi2020tight}.

For the complementary minimization objectives, the obstruction is already
visible when the optimum value is zero.  Every approximation with a finite
factor must then find a zero-valued solution exactly.  Deciding whether at most
$k$ sets cover every element, or whether an assignment of Hamming weight at
most $k$ satisfies every constraint, is $\Wtwo$-hard by the standard hardness
of \textsc{Set Cover} and its monotone-SAT encoding.  Consequently, unless
$\FPT=\Wtwo$, none of the three unrestricted problems admits a finite-factor
FPT approximation parameterized by $k$ (\cref{cor:unrestricted-hardness}).
All objective-value hardness proofs in this paper distinguish zero optimum
from positive optimum.  They do not establish a multiplicative gap under
the promise that the optimum is positive.
This sharp contrast, together with the inapproximability of the unrestricted
problems, shows that obtaining any meaningful finite-factor approximation
requires imposing additional structure. This motivates our study of natural
restrictions of the incidence graph under which the residual-minimization
objectives become approximable.

Alongside approximating the minimum uncovered or unsatisfied weight while
using at most $k$ selectable objects, we study a complementary question. Given a target attained by a
solution using at most $k$ selectable objects, can an FPT algorithm preserve
that target while using at most $g(k,d)$ objects, where $g$ is a computable
function of the parameters and $g(k,d)\geq k$? In this setting, the
maximization and residual-minimization formulations are exactly equivalent:
leaving uncovered or unsatisfied weight at most $t$ is the same as covering
or satisfying weight at least the corresponding complementary target.
Thus, unlike multiplicative approximation of the objective value, there is
no distinction between the two viewpoints once the target is fixed and only
the cardinality bound is approximated.  Consequently, any target-preserving
result for one formulation immediately transfers to the other. For coverage, this is the weighted \textsc{Partial Set Cover} problem: given
a prescribed coverage threshold, minimize the number of sets whose union has
total weight at least that threshold. We return to this question after introducing the structural restrictions
considered in the paper and presenting our results on approximating the
uncovered or unsatisfied weight.

\paragraph*{Structural restrictions.}
We study five principal restrictions on the incidence graph
$G=(A\uplus B,E)$.  For \MinUnCoverage{}, $A=\mathcal S$ and $B=U$, with
$Su\in E$ if and only if $u\in S$.  For \MinUnsatSAT{} and
\MinUnsatCSP{}, $A$ is the set of variables and $B$ is the set of clauses or
constraints, with $xC\in E$ if and only if $x$ occurs in $C$.  Thus, the
cardinality bound $k$ restricts the choices made on side $A$, while the vertices whose
weights contribute to the residual lie on side $B$.
The first two restrictions bound degrees on opposite sides of the incidence
graph.  Bounded arity means that every clause or constraint contains at most
$d$ variables; for \MinUnCoverage{}, the corresponding condition is bounded
element frequency, meaning that every ground element belongs to at most $d$
sets.  Conversely, bounded variable frequency means that every variable
occurs in at most $d$ clauses or constraints; on the coverage side, this
corresponds to bounded set size.  These two restrictions are incomparable.

Beyond these two one-sided degree bounds, we consider three increasingly
general restrictions on the incidence graph.  The first is
$K_{d,d}$-freeness: the incidence graph contains no complete bipartite
subgraph with $d$ vertices on each side.  A degree bound of $d$ on either
side implies $K_{d+1,d+1}$-freeness.  Next, we consider bounded semi-ladder
index, a notion introduced by Fabia\'nski et al.
\cite{fabianski2019progressive}.  A semi-ladder of order $\ell$ consists of
vertices $a_1,\ldots,a_\ell$ on one side and $b_1,\ldots,b_\ell$ on the other
such that $a_i b_i$ is a nonedge and $a_i b_j$ is an edge whenever
$j<i$ (\cref{def:semiladder}).
The semi-ladder index is the maximum order of such a pattern.
Every $K_{d,d}$-free graph has semi-ladder index less than $2d$, while set
systems of semi-ladder index two can contain arbitrarily large bicliques.  For
example, a disjoint union of complete bipartite graphs has semi-ladder index
at most two and may contain a $K_{n,n}$ for arbitrarily large $n$.
Thus bounded semi-ladder index is strictly more general than
biclique-freeness.  The final restriction is bounded VC dimension.
The neighborhoods of the vertices in $A$ form the set system $  \{N_G(a):a\in A\}\subseteq 2^B$. A set $X\subseteq B$ is \emph{shattered} if, for every $Y\subseteq X$, there
exists a vertex $a\in A$ such that $N_G(a)\cap X=Y$.  The VC dimension is the
maximum cardinality of a shattered set (\cref{def:vc-dimension}).  Every
shattered set of size $\ell$ yields a semi-ladder of order $\ell$
(\cref{obs:vc-semiladder}), so bounded semi-ladder index implies bounded VC
dimension.  The converse fails: half-graphs have VC dimension one but
unbounded semi-ladder index; a half-graph of order $\ell$ has vertices
$a_1,\ldots,a_\ell$ and $b_1,\ldots,b_\ell$, with $a_i b_j$ an edge exactly
when $j<i$.

\paragraph*{Our results.}
We first present our results on approximating the residual weight while
respecting the cardinality bound, organized according to
the hierarchy of incidence restrictions described above.  We then turn to
approximating the cardinality bound while preserving a prescribed upper bound
on the residual weight.

\paragraph*{Bounded arity and ground-element frequency.}
For every $d\geq 1$, we give a deterministic efficient parameterized
approximation scheme (EPAS) for weighted
arity-$d$ \MinUnsatCSP{} parameterized by $k+d$
(\cref{thm:arity-csp}).  For $0<\eps\leq 1$, its running time is
\[
  (dk/\eps)^{O(k)}N^{O(1)},
\]
where $N$ denotes the input encoding size.  Since CNF formulas are a special
case of Boolean CSP, the same result applies to weighted arity-$d$
\MinUnsatSAT{}.  Moreover, encoding each ground element as a monotone clause
of the same weight gives a deterministic EPAS for weighted
\MinUnCoverage{} when every ground element belongs to at most $d$ sets
(\cref{cor:arity-special-cases}).  Skowron and Faliszewski previously gave a
randomized FPT approximation scheme for unit weights; our algorithm is
deterministic and also handles positive rational weights
\cite{skowron2017boundedfrequencies}.

The algorithm compares the all-zero assignment with an optimum assignment at
each recursive step.  If the all-zero assignment is not already a
$(1+\eps)$-approximation, then some variable set to one by the optimum occurs
in constraints of sufficiently large total weight that the all-zero
assignment violates.  Bounded arity limits the number of variables with this
property, allowing the algorithm to branch over them in FPT time.

\paragraph*{Bounded variable frequency and set size.}
This pair of corresponding one-sided degree restrictions yields a sharp
separation between coverage and general satisfaction problems.  In the
monotone-CNF encoding of \MinUnCoverage{}, each set $S$ is represented
by a variable $x_S$, which occurs in one clause for every element of $S$.
Thus, a bound of $d$ on set size is exactly a bound of $d$ on variable
frequency.
Bl\"aser's randomized color-coding algorithm gives an exact FPT algorithm for unweighted
\textsc{Partial Set Cover}~\cite{DBLP:journals/ipl/Blaser03}; Bonnet et al.
gave an alternative deterministic FPT algorithm for
\textsc{Maximum $k$-Set Cover} based on greediness-for-parameterization
\cite{bonnet2016maxksetcover}.  We adapt Bl\"aser's color-coding framework to
positive rational element weights and show that weighted \MinUnCoverage{} is
exactly solvable in
\[
  2^{O(kd)}N^{O(1)}
\]
time when every set has size at most $d$
(\cref{prop:bounded-set-size-coverage}).  Our extension assigns each covered
element (at most $kd$ many) to one selected set, producing at most $k$
pairwise disjoint subsets, each of size at most $d$.  A $kd$-perfect hash
family contains a coloring that is
injective on all elements covered by an optimum solution, and a dynamic
program over subsets of colors finds a maximum-weight collection of such
subsets while remembering the original set containing each one.  Selecting
those original sets recovers an optimum coverage solution.  Since this proof
is a weighted deterministic adaptation of Bl\"aser's color-coding argument,
rather than one of our main
structural techniques, we defer it to \cref{sec:bounded-set-size}.  In contrast,
we show that unless $\FPT=\Wone$, neither \MinUnsatSAT{} nor
\MinUnsatCSP{} admits a finite-factor FPT approximation parameterized by
$k+d$, even on unweighted instances in which every variable occurs in at most
$d$ clauses or constraints (\cref{thm:frequency-sat-hard}).  Under the
Exponential Time Hypothesis (ETH), no
such approximation runs in $f(k,d)N^{o(\sqrt{k+d})}$ time for any computable
function $f$.

\paragraph*{$K_{d,d}$-free graphs and bounded semi-ladder index.}
We now move from one-sided degree bounds to restrictions on the entire
incidence graph.  For weighted \MinUnCoverage{} on incidence graphs of
semi-ladder index at most $d$, we give a deterministic EPAS parameterized by
$k+d$ (\cref{thm:semiladder-uncoverage}), with running time
\[
  \left(\frac{dk}{\eps}\right)^{
    O\left(
      \frac{d^2k}{\eps}\log\frac{2k}{\eps}
    \right)
  }N^{O(1)}.
\]
The algorithm adapts the skeleton-based branching framework and deterministic
weighted-net construction of Kundu et al.~\cite{kundu2026fixedbudget}, which in turn
builds on Guillemot's FPT algorithm for \textsc{Set Cover} on semi-ladder-free
hypergraphs~\cite{guillemot2025parameterized}.  The skeletons, eligible
families, forced-coverage intersections, weighted nets, and semi-ladder depth
argument are inherited from that framework.  Our additional step is the
residual-weight case distinction and averaging argument described next, which gives
the new approximation guarantee.  The algorithm maintains $k$
skeletons $M_1,\ldots,M_k$, where $M_i$ records elements that the $i$th chosen
set must contain.  The sets containing $M_i$ remain eligible for that
coordinate, and their intersection is already guaranteed to be covered.  Let
$W$ be the elements not guaranteed to be covered at any coordinate.  If
$w(W)>(1+\eps)\OPT_{\mathrm{miss}}(I)$, an averaging argument shows that some
optimum set $O_i$ covers weight greater than $\delta w(W)$, where
$\delta=\eps/(k(1+\eps))$.  A weighted
$\delta$-net then contains an element $u\in O_i\cap W$; since the algorithm
tries every pair $(i,u)$, one recursive branch adds $u$ to the corresponding
skeleton and remains consistent with the fixed optimum.  Each element added
to one skeleton produces a new semi-ladder rung, so every skeleton has size at
most $d$ and the recursion depth is at most $kd$.  When $w(W)$ is small,
arbitrary eligible sets already give the required approximation.

By \cref{obs:kdd-semiladder}, every $K_{d,d}$-free graph has semi-ladder
index less than $2d$.  Consequently, \cref{thm:semiladder-uncoverage} also
gives an EPAS for weighted $K_{d,d}$-free \MinUnCoverage{}.  For the unit-weight case, we additionally give a
specialized algorithm with the improved running time
$(dk/\eps)^{O(dk)}N^{O(1)}$ (\cref{thm:kdd-coverage}).

These graph-based positive results rely on the monotonicity of coverage and
do not extend to general CNF formulas.  Unless $\FPT=\Wone$,
\MinUnsatSAT{} and \MinUnsatCSP{} admit no finite-factor FPT approximation
even on unweighted $K_{3,3}$-free instances and, separately, on unweighted
instances of semi-ladder index at most four
(\cref{thm:kdd-sat-hard}).  Under ETH, no such approximation runs in
$f(k,d)N^{o(k+d)}$ time for any computable function $f$.

\paragraph*{Bounded VC dimension.}
We compare the preceding graph restrictions with bounded VC dimension.
Badanidiyuru et al.\ gave an EPAS for \textsc{Maximum Coverage} on set
systems of bounded VC dimension~\cite{badanidiyuru2012lowdimensional}.
The corresponding residual objectives behave differently.  Bringmann et al.\
proved that \textsc{Hitting Set} is
$\Wone$-hard even when both its primal and dual VC dimensions are at most
two~\cite{bringmann2016hittingset}.  Dualizing their instances and using the
fact that any finite-factor approximation must distinguish optimum zero from
positive optimum (\cref{prop:zero-barrier}) shows that, unless $\FPT=\Wone$, none of
\MinUnCoverage{}, \MinUnsatSAT{}, and \MinUnsatCSP{} admits a
finite-factor FPT approximation on unweighted instances of VC dimension at
most two (\cref{thm:vc-dimension-hardness}).  Thus, bounded semi-ladder index
suffices for approximating the coverage residual, whereas bounded VC
dimension alone does not.

\Cref{tab:dichotomy} summarizes the FPT approximability landscape for the residual-minimization objective (the
uncovered or unsatisfied weight) under the cardinality bound $k$, with
parameter $k+d$, for each incidence restriction considered in the paper.
Positive results are for positive rational weights unless stated otherwise, whereas all hardness
results already hold for unweighted instances.
Each hardness entry rules out every finite-factor FPT approximation
parameterized by $k+d$, unless $\FPT=\Wone$.

\begin{table}[ht]
  \centering
  \scriptsize
  \setlength{\tabcolsep}{2.5pt}
  \renewcommand{\arraystretch}{1.12}
  \begin{tabularx}{\textwidth}{@{}
    >{\raggedright\arraybackslash}p{0.26\textwidth}
    >{\raggedright\arraybackslash}p{0.24\textwidth}
    >{\centering\arraybackslash}p{0.19\textwidth}
    >{\raggedright\arraybackslash}X@{}}
    \toprule
    \textbf{Incidence restriction}
      & \textbf{\MinUnCoverage{}}
      & \textbf{\MinUnCoverage{} running time}
      & \textbf{\MinUnsatSAT{}/\MinUnsatCSP{}} \\
    \midrule
    clause or constraint arity $\leq d$;\newline
    ground-element frequency $\leq d$
      & \textbf{Consequence:} deterministic EPAS for weighted instances
        (\cref{cor:arity-special-cases});\newline
        \textbf{External:} unit-weight randomized EPAS
        \cite{skowron2017boundedfrequencies}
      & $(dk/\eps)^{O(k)}N^{O(1)}$
      & \textbf{This paper:} deterministic EPAS for weighted instances
        (\cref{thm:arity-csp});\newline $(dk/\eps)^{O(k)}N^{O(1)}$ \\
    \midrule
    variable frequency $\leq d$;\newline
    set size $\leq d$
      & \textbf{Adaptation:} exact FPT algorithm for weighted instances,
        extending~\cite{DBLP:journals/ipl/Blaser03}
        (\cref{prop:bounded-set-size-coverage})
      & $2^{O(kd)}N^{O(1)}$
      & \textbf{This paper:} no finite-factor FPT approximation
        (\cref{thm:frequency-sat-hard}) \\
    \midrule
    $K_{d,d}$-free incidence graph
      & \textbf{Consequence:} deterministic EPAS for weighted instances
        (\cref{thm:semiladder-uncoverage});\newline
        \textbf{This paper:} improved parameter dependence for unit weights
        (\cref{thm:kdd-coverage})
      & weighted: $(dk/\eps)^{O(L)}N^{O(1)}$;\newline
        unit: $(dk/\eps)^{O(dk)}N^{O(1)}$
      & \textbf{This paper:} no finite-factor FPT approximation on unweighted
        $K_{3,3}$-free instances
        (\cref{thm:kdd-sat-hard}) \\
    \midrule
    semi-ladder index $\leq d$
      & \textbf{This paper:} deterministic EPAS for weighted instances
        (\cref{thm:semiladder-uncoverage})
      & $(dk/\eps)^{O(L)}N^{O(1)}$
      & \textbf{This paper:} no finite-factor FPT approximation on unweighted
        instances of semi-ladder index at most four
        (\cref{thm:kdd-sat-hard}) \\
    \midrule
    VC dimension $\leq d$
      & \textbf{Consequence:} no finite-factor FPT approximation,
        even for $d=2$ (\cref{thm:vc-dimension-hardness},
        from~\cite{bringmann2016hittingset})
      & ---
      & \textbf{Consequence:} no finite-factor FPT approximation,
        even for $d=2$ (\cref{thm:vc-dimension-hardness},
        from~\cite{bringmann2016hittingset}) \\
    \bottomrule
  \end{tabularx}
  \caption{Residual-objective results for parameter $k+d$.  ``This paper''
  marks results proved here; ``Consequence'' marks an immediate corollary or
  specialization; ``Adaptation'' marks a weighted deterministic extension
  of a cited algorithm; and ``External'' marks cited work.
  All hardness entries use the zero-optimum barrier and do not give a gap
  between two positive optimum values.
  Here $N$ is the input size and
  $L=(d^2k/\eps)\log(2k/\eps)$.  In the two one-sided degree rows, the first
  line gives the SAT/CSP restriction and the second line gives the corresponding
  coverage restriction.}
  \label{tab:dichotomy}
\end{table}

\paragraph*{Approximating the cardinality bound.}
We finally consider relaxing the cardinality bound while preserving a
prescribed upper bound on the uncovered or unsatisfied weight.  As discussed above, the maximization and
residual-minimization formulations are equivalent once the target is fixed.
On the coverage side, Jain et al.~\cite{jain2023parameterized} showed that,
on $K_{d,d}$-free incidence graphs, a target attainable with $k$ sets can be
attained in FPT time with at most $k+1$ sets.  Kundu et al.~\cite{kundu2026fixedbudget}
extended this additive-one
guarantee to weighted \textsc{Partial Set Cover} on incidence graphs of bounded
semi-ladder index.  Their result therefore also
applies to the biclique-free and one-sided degree-bounded coverage classes
considered here.  For bounded set size, our exact FPT algorithm attains the
target without increasing the cardinality bound.  In contrast, bounded VC dimension
alone does not permit an additive-one guarantee: for every fixed
$\delta>0$, \textsc{Partial Set Cover} admits no parameterized
$(2-\delta)$-approximation of the cardinality bound, unless $\FPT=\Wone$,
even on set systems of VC dimension seven~\cite{kundu2026fixedbudget}.

For \MinUnsatSAT{} and \MinUnsatCSP{}, the separation is stronger: all of our
hardness results below already hold for unweighted instances.  Unless
$\FPT=\Wone$, for every computable function $g(k,d)\geq k$, no FPT algorithm
can preserve a prescribed upper bound on the unsatisfied weight while
replacing the cardinality bound $k$ by $g(k,d)$.  This holds for clause or
constraint arity three (\cref{thm:cardinality-arity}), under bounded variable
frequency (\cref{thm:cardinality-frequency}), and consequently on
$K_{3,3}$-free incidence graphs, on incidence graphs of semi-ladder index at
most four, and on incidence systems of VC dimension at most two
(\cref{cor:cardinality-other-restrictions}).  These reductions establish a
stronger gap than the failure of any particular approximation factor: a
yes-instance has an assignment of Hamming weight at most $k$ attaining the
prescribed bound, whereas in a no-instance no assignment, regardless of its
Hamming weight, attains it.  Equivalently, let \textsc{CC-Max-SAT} denote the
problem of maximizing the total weight of satisfied clauses subject to a bound
on the Hamming weight.  Our results give a negative answer to its corresponding
``optimizing $k$'' question under all the structural restrictions considered
here: even an arbitrary computable relaxation of the cardinality bound is
impossible.  This contrasts with the additive-one FPT guarantee for
\textsc{Partial Set Cover}, proved by Jain et al.~\cite{jain2023parameterized} on
$K_{d,d}$-free incidence graphs and extended by
Kundu et al.~\cite{kundu2026fixedbudget} to weighted instances of bounded semi-ladder index. Our lower bounds thereby complete the
corresponding picture on the satisfaction side of the target-preserving cardinality-approximation question initiated by Jain et al.

\Cref{tab:cardinality-bound} summarizes the results for relaxing the
cardinality bound while preserving the objective target.  The coverage column
states how many sets suffice when $k$ sets attain the target.  The SAT/CSP
column records the corresponding cardinality-approximation hardness.
Every hardness entry in the SAT/CSP column holds unless $\FPT=\Wone$.

\begin{table}[H]
\centering
\small
\setlength{\tabcolsep}{3pt}
\renewcommand{\arraystretch}{1.12}
\begin{tabularx}{\textwidth}{
    @{}p{0.25\textwidth}
       p{0.34\textwidth}
       X@{}}
\toprule
\textbf{Incidence restriction}
&
\textbf{\MinUnCoverage{}}
&
\textbf{\MinUnsatSAT{}/\MinUnsatCSP{}}
\\
\midrule

Clause or constraint arity $\leq d$;\newline
ground-element frequency $\leq d$
&
Target preserved with at most $k+1$ sets in FPT time parameterized by $k+d$;
weighted instances
\cite{kundu2026fixedbudget, jain2023parameterized}
&
No computable cardinality relaxation in FPT time, already for
$d=3$
(\cref{thm:cardinality-arity,cor:cardinality-other-restrictions})
\\
\midrule

Variable frequency $\leq d$;\newline
set size $\leq d$
&
Exact FPT algorithm parameterized by $k+d$
(\cref{prop:bounded-set-size-coverage})
&
No computable cardinality relaxation in FPT time parameterized
by $k+d$
(\cref{thm:cardinality-frequency,cor:cardinality-other-restrictions})
\\
\midrule

$K_{d,d}$-free incidence graph
&
Target preserved with at most $k+1$ sets in FPT time parameterized by $k+d$;
weighted instances
\cite{kundu2026fixedbudget,jain2023parameterized}
&
No computable cardinality relaxation in FPT time, already on
$K_{3,3}$-free incidence graphs
(\cref{cor:cardinality-other-restrictions})
\\
\midrule

Semi-ladder index $\leq d$
&
Target preserved with at most $k+1$ sets in FPT time parameterized by $k+d$;
weighted instances
\cite{kundu2026fixedbudget}
&
No computable cardinality relaxation in FPT time, already for
semi-ladder index four
(\cref{cor:cardinality-other-restrictions})
\\
\midrule

VC dimension $\leq d$
&
No FPT approximation within $2-\delta$, for any fixed $\delta>0$, even for
$d=7$~\cite{kundu2026fixedbudget}
&
No computable cardinality relaxation in FPT time, already for
$d=2$ (\cref{cor:cardinality-other-restrictions})
\\
\bottomrule
\end{tabularx}
\caption{Results for relaxing the cardinality bound while preserving the
bound on uncovered or unsatisfied weight.  The parameter is $k$ unless stated
otherwise; all SAT/CSP hardness results hold for unweighted instances.  In the
first two rows, the restrictions before and after the semicolon refer to
SAT/CSP and coverage, respectively.}
\label{tab:cardinality-bound}
\end{table}

\paragraph*{Applications of the \MinUnsatCSP{} formulation.}
The CSP formulation also captures cardinality-constrained selection problems
in which covering some elements incurs a penalty.  Such objectives are
nonmonotone and therefore fall outside standard coverage maximization.  We
illustrate this with two related examples.

An instance of \textsc{Positive--Negative Partial Set Cover} consists of
disjoint sets $P$ and $N$ of positive and negative elements, respectively,
and a family $\mathcal S\subseteq 2^{P\uplus N}$.  The objective is to choose
a subfamily $\mathcal T\subseteq\mathcal S$ minimizing the number of uncovered
positive elements plus the number of covered negative elements.  We consider
the weighted, cardinality-constrained version, in which all elements have
positive rational weights and $|\mathcal T|\leq k$.

An instance of \textsc{Red--Blue Set Cover} consists of disjoint sets $R$ and
$B$ of red and blue elements, respectively, and a family
$\mathcal S\subseteq 2^{R\uplus B}$.  Its objective is to cover every blue
element while minimizing the number of covered red elements.  In the weighted,
cardinality-constrained version considered here, red elements have positive
rational weights and at most $k$ sets may be selected.

Carr et al.\ introduced \textsc{Red--Blue Set
Cover}~\cite{carr2000redblue}.  Peleg subsequently gave a
$2\sqrt{m\log |B|}$-approximation, where $m=|\mathcal S|$
\cite{peleg2007redblue}.  Chlamt\'a\v{c} et al.\ obtained an
$O(m^{1/3}\log^{4/3}|R|\log |B|)$-approximation
\cite{chlamtac2023redblue}.  Miettinen introduced
\textsc{Positive--Negative Partial Set Cover} and gave polynomial-time
reductions in both directions between it and \textsc{Red--Blue Set Cover},
thereby transferring approximation guarantees and hardness results between
the two problems.  For the original unweighted problem without a cardinality
bound, he also proved that, for every $\eps>0$, no polynomial-time
$2^{(\log |P|)^{1-\eps}}$-approximation exists unless
$\mathsf{P}=\mathsf{NP}$~\cite{miettinen2008positive}.

We next encode the weighted, cardinality-constrained variants as
\MinUnsatCSP{} instances.  For every set $S\in\mathcal S$, introduce a
variable $x_S$, where $x_S=1$ means that $S$ is selected.  A positive element
$p$ is represented by the constraint
\[
  C_p=\bigvee_{S\ni p}x_S,
\]
which is unsatisfied precisely when $p$ is uncovered.  A negative element
$n$ is represented by
\[
  C_n=\bigwedge_{S\ni n}\neg x_S,
\]
which is unsatisfied precisely when $n$ is covered.  Give each constraint the
weight of its corresponding element.  Thus, the total weight
of the unsatisfied constraints is exactly the total weight of the uncovered
positive elements and the covered negative elements, that is, the objective
value of the corresponding \textsc{Positive--Negative Partial Set Cover}
solution.

For the weighted, cardinality-constrained \textsc{Red--Blue Set Cover}
encoding, in which blue elements play the role of positive elements and red
elements play the role of negative elements, fix the requested accuracy
$\eps>0$, let $W_R$ be the total weight
of the red elements, and give every blue constraint a rational weight
$M>(1+\eps)W_R$; red constraints retain their original weights.  Suppose that
all blue elements can be covered with at most $k$ sets.  The optimum CSP value
is then at most $W_R$, whereas any assignment leaving a blue element uncovered
has value at least $M$.  Consequently, both an optimum solution and every
$(1+\eps)$-approximate solution cover all blue elements, and the CSP objective
on such solutions is exactly the covered red weight.  This gives the desired
$(1+\eps)$-approximation for every feasible instance.  Because selecting an
additional set can violate a negative-element constraint, these encodings
require general \MinUnsatCSP{} rather than \MinUnCoverage{}.

In these encodings, bounded arity has a direct interpretation in the original
problems.  For \textsc{Positive--Negative Partial Set Cover} and
\textsc{Red--Blue Set Cover}, the constraint associated with an element has
arity equal to the number of sets containing that element.  Hence, arity at
most $d$ means that every ground element belongs to at most $d$ sets.  Our
bounded-arity EPAS therefore applies directly under this restriction.

Bounded set size alone gives a different picture for the nonmonotone
red--blue objective.  For the unweighted decision version, called
\textsc{Generalized Red--Blue Set Cover}, Ashok et al.\ proved
$\Wone$-hardness parameterized by the number of selected sets and the red
budget even when every set has size three and contains at least two red
elements.  They also gave a $2^{O(dk)}N^{O(1)}$-time exact algorithm when
every set contains at most one red element and at most $d$ blue
elements~\cite{ashok2017redblue}.  This contrast explains why the exact
bounded-set-size result in \cref{prop:bounded-set-size-coverage} is specific
to the monotone coverage objective.

\paragraph*{Related work.}
The survey of Feldmann et al. calls the unit-weight version of our coverage
objective \textsc{Min $k$-Uncovered},
observes that it is not approximable on unrestricted set systems, and records
the randomized bounded-frequency scheme as Theorem~21
\cite[Section~4.2.3]{feldmann2020parameterizedsurvey}.  In the original
analysis, Skowron and Faliszewski call the problem \textsc{MinNonCovered} and prove the
scheme in their Theorem~4.5~\cite{skowron2017boundedfrequencies}.  They also
record in Table~1 that exact optimization is $\Wone$-complete for every fixed
upper frequency bound at least two when parameterized by $k$.  When the permitted
number $\ell$ of uncovered elements is the parameter, the problem is
para-NP-complete even under bounded element frequency; for unrestricted set
systems, it is $\Wtwo$-complete for the joint parameter $(k,\ell)$.

For the graph case, which has ground-element frequency two, \textsc{Maximum Partial
Vertex Cover} is $\Wone$-complete when parameterized by $k$
\cite{guo2007vertexcover}.  Damaschke studied
the equivalent problem of selecting at most $k$ vertices while leaving at
most $\ell$ edges uncovered.  He gave exact FPT branching algorithms
parameterized by $k+\ell$~\cite{damaschke2009pareto}.  These algorithms use
the permitted number of uncovered elements as an additional parameter and do
not give an approximation parameterized only by $k$.

The semi-ladder class and terminology originate with Fabia\'nski et
al.~\cite{fabianski2019progressive}.  Guillemot subsequently studied \textsc{Set Cover}
on precisely this class, adapting ideas of Langerman and Morin
\cite{langerman2005covering} to obtain FPT
and kernelization algorithms together with a matching compression lower bound
\cite{guillemot2025parameterized}.  This exact-cover work is a
structural predecessor of our semi-ladder result.  Our objective is different:
we permit uncovered elements and seek a multiplicative approximation of their
minimum total weight without taking that weight as a parameter.
Kundu et al.~\cite{kundu2026fixedbudget} subsequently identified a related boundary
for weighted \textsc{Partial Set Cover}.  They showed that bounded semi-ladder index
permits an FPT algorithm to preserve a coverage target with $k+1$ sets whenever
$k$ sets suffice.  Conversely, for every fixed $\delta>0$, they ruled out a
parameterized $(2-\delta)$-approximation on set systems of VC dimension seven
unless $\FPT=\Wone$.  They also developed the skeleton framework, branching
idea, and weighted-net bounds used in our objective-value approximation; their
construction explicitly builds on Guillemot's FPT routine
\cite{guillemot2025parameterized}.

For broader background on approximation in parameterized complexity, we refer
to Feldmann et al.~\cite{feldmann2020parameterizedsurvey}.  Jain et al.\ gave
approximation schemes for cardinality-constrained \textsc{Max $k$-Weight SAT}
and \textsc{Maximum Coverage} on biclique-free incidence systems
\cite{jain2023parameterized}.  Manurangsi subsequently obtained an improved
deterministic scheme and an approximate kernel for
\textsc{Max $k$-Weight SAT}~\cite{manurangsi2025improved}.  Recent work also
gives reductions between satisfaction and coverage in the presence of global
and matroid constraints~\cite{inamdar2024satisfiability}.  Inamdar et al.\ further
obtained parameterized approximation
schemes for connectivity-constrained \textsc{Maximum Coverage} on
$K_{d,d}$-free incidence graphs~\cite{inamdar2026connected}.  Those results
maximize the number of satisfied clauses or covered elements and therefore do
not directly control the small complementary objective studied here.  CSPs with global
cardinality constraints have also been studied through semidefinite
programming hierarchies~\cite{raghavendra2012global}.

Finally, our formulation should not be confused with \textsc{Boolean MinCSP} as studied
in the constraint-deletion literature
\cite{bonnet2016mincsp,kim2025flowaugmentation}.  There, the parameter bounds
the number or total weight of constraints that may be violated, so the input
is required to be close to satisfiable.  Here the parameter $k$ instead bounds
the Hamming weight of the assignment; the number of violated constraints is
unbounded and is the objective being minimized.

\paragraph*{Organization.}
\Cref{sec:preliminaries} introduces the problems and the relevant incidence parameters.
\Cref{sec:bounded-arity} presents an EPAS for bounded-arity \MinUnsatCSP{}, while
\cref{sec:kdd-hardness} establishes hardness results for \MinUnsatSAT{} and
\MinUnsatCSP{} under biclique-freeness, bounded semi-ladder index, and bounded
variable frequency. In contrast, \cref{sec:semiladder-uncoverage} develops an
EPAS for \MinUnCoverage{} on instances of bounded semi-ladder index, including
its biclique-free consequence. \Cref{sec:cardinality-augmentation}
investigates the alternative approximation of the cardinality
bound and completes the landscape for all three problems under each of the
considered restrictions. \Cref{sec:conclusion} summarizes the resulting
landscape. Finally, the appendices contain a specialized EPAS for
unit-weight biclique-free instances and an exact algorithm for weighted
\MinUnCoverage{} with bounded set size.

%% file: sections/preliminaries.tex
\section{Preliminaries}
\label{sec:preliminaries}

A Boolean constraint is a pair
$C=(\operatorname{scope}(C),R_C)$, where the scope is an ordered tuple of
distinct variables and $R_C\subseteq\{0,1\}^{|\operatorname{scope}(C)|}$ is a
relation.  The assignment $\tau$ satisfies $C$ when its restriction to the
ordered scope, viewed as a tuple in $\{0,1\}^{|\operatorname{scope}(C)|}$,
belongs to $R_C$.  A weighted Boolean CSP instance is a triple
$(X,\mathcal C,w)$ consisting of variables, a collection of constraints, and
a weight function $w\colon\mathcal C\to\mathbb Q_{>0}$.
The arity of the instance is the maximum scope size.  A literal is a Boolean
variable or its negation.  A clause is a disjunction of literals, and a CNF
formula is a conjunction of clauses.  We remove repeated literals and discard
tautological clauses, which does not change the unsatisfied weight of any
assignment.
We represent each constraint relation extensionally by listing either its
satisfying tuples or its falsifying tuples, together with a bit indicating
which list is used.  Thus, a clause is represented by its unique falsifying
tuple, making SAT a syntactic special case of Boolean CSP.  An arity-$d$
relation has at most $2^d$ tuples, so this representation is compatible with
the parameterized running times below.

\begin{definition}[\MinUnsatCSP{} and \MinUnsatSAT]
Given a weighted Boolean CSP instance $(X,\mathcal C,w)$ and a nonnegative
integer $k$, the objective is to find an assignment $\tau$ minimizing
\[
  \unsat_{\mathcal C,w}(\tau)
  = \sum_{\substack{C\in\mathcal C\\
                     \tau\text{ does not satisfy }C}}w(C).
\]
The assignment must have Hamming weight at most $k$, where its Hamming weight
is $|\{x\in X:\tau(x)=1\}|$.
For $\mathcal D\subseteq\mathcal C$, write
$w(\mathcal D)=\sum_{C\in\mathcal D}w(C)$.  If
$I=(X,\mathcal C,w)$, we also write $\unsat_I(\tau)$ for
$\unsat_{\mathcal C,w}(\tau)$.
When every constraint is a clause of a CNF formula, the problem is
\MinUnsatSAT{}.  An instance is unweighted when every constraint or clause has
weight one.
\end{definition}

\begin{definition}[\MinUnCoverage]
An instance is a tuple $I=(U,\mathcal S,w,k)$, where $U$ is a set of ground
elements, $\mathcal S\subseteq 2^U$ is a family of sets,
$w\colon U\to\mathbb Q_{>0}$ is a weight function, and $k$ is a nonnegative
integer.
The objective is to find a subfamily $\mathcal A\subseteq\mathcal S$ with
$|\mathcal A|\le k$ minimizing
\[
  \miss_{\mathcal S,w}(\mathcal A)
  = \sum_{u\in U\setminus\bigcup_{S\in\mathcal A}S}w(u).
\]
For $Y\subseteq U$, write $w(Y)=\sum_{u\in Y}w(u)$.  We also write
$\cov_{\mathcal S,w}(\mathcal A)=w(U)-\miss_{\mathcal S,w}(\mathcal A)$
for the total weight of the covered elements.  When all ground elements have
weight one, the instance is unweighted and we omit $w$ from the notation.
\end{definition}

\begin{definition}[Incidence graph]
For a CSP instance $I=(X,\mathcal C,w)$, the incidence graph $G_I$ is the
bipartite graph with parts $X$ and $\mathcal C$.  It contains an edge $xC$ if
and only if variable $x$ occurs in the scope of constraint $C$.  For SAT, the
two parts are the variables and clauses, with an edge recording that a variable
occurs in a clause.  For a coverage instance $I=(U,\mathcal S,w,k)$, the two
parts are the sets $\mathcal S$ and the ground elements $U$.  It
contains an edge $uS$ if and only if $u\in S$.
\end{definition}

\begin{definition}[VC dimension~\cite{vapnik1971uniform}]
\label{def:vc-dimension}
Let $(U,\mathcal F)$ be a set system, where $\mathcal F\subseteq 2^U$.  A set
$X\subseteq U$ is \emph{shattered} by $\mathcal F$ if
\[
  \{F\cap X:F\in\mathcal F\}=2^X.
\]
The \emph{VC dimension} of $(U,\mathcal F)$ is the maximum cardinality of a
set shattered by $\mathcal F$; it is also called the \emph{primal VC
dimension}.  The \emph{dual set system} is $(\mathcal F,\mathcal F^*)$, where
\[
  \mathcal F^*
  :=\bigl\{\{F\in\mathcal F:u\in F\}:u\in U\bigr\}.
\]
Its VC dimension is the \emph{dual VC dimension} of $(U,\mathcal F)$.

For a bipartite incidence graph $G=(A\uplus B,E)$, where $A$ contains the
selectable objects, let
\[
  \mathcal N_A(G):=\{N_G(a):a\in A\}\subseteq 2^B,
  \qquad
  \mathcal N_B(G):=\{N_G(b):b\in B\}\subseteq 2^A.
\]
The primal and dual VC dimensions of $G$ are the VC dimensions of
$(B,\mathcal N_A(G))$ and $(A,\mathcal N_B(G))$, respectively.  Unless stated
otherwise, the VC dimension of $G$, and hence of the corresponding instance,
means its primal VC dimension.
\end{definition}

\begin{definition}[Semi-ladder index~\cite{fabianski2019progressive}]
\label{def:semiladder}
Let $G=(A\uplus B,E)$ be a bipartite graph.  A \emph{semi-ladder of order
$\ell$} consists of vertices $a_1,\ldots,a_\ell\in A$ and
$b_1,\ldots,b_\ell\in B$ such that
\[
  a_i b_i\notin E \quad\text{for every $i\in[\ell]$},
  \qquad
  a_i b_j\in E \quad\text{whenever $j<i$}.
\]
There is no condition on $a_i b_j$ when $j>i$.  The \emph{semi-ladder index}
of $G$ is the maximum order of a semi-ladder in $G$.
\end{definition}

An instance is $K_{d,d}$-free if its incidence graph contains no $K_{d,d}$ as
a subgraph.

\begin{observation}
\label{obs:kdd-semiladder}
For every $d\ge1$, every $K_{d,d}$-free bipartite graph has semi-ladder index
less than $2d$.
\end{observation}

\begin{proof}
If $a_1,\ldots,a_{2d}$ and $b_1,\ldots,b_{2d}$ form a semi-ladder, then
$a_{d+1},\ldots,a_{2d}$ and $b_1,\ldots,b_d$ induce all edges of a
$K_{d,d}$, since $j\le d<i$ for every selected pair.  Thus a
$K_{d,d}$-free graph has no semi-ladder of order $2d$.  Since every prefix of
a semi-ladder is itself a semi-ladder, it has no semi-ladder of order at least
$2d$, and hence its semi-ladder index is less than $2d$.
\end{proof}

\begin{observation}[Kundu et al.]
\label{obs:vc-semiladder}
The VC dimension of a bipartite incidence graph is at most its semi-ladder
index~\cite[Lemma~8.6]{kundu2026fixedbudget}.
\end{observation}

\begin{proof}
Let $X=\{b_1,\ldots,b_\ell\}$ be shattered.  For each $i\in[\ell]$, choose
$a_i$ whose neighborhood satisfies
$N(a_i)\cap X=\{b_1,\ldots,b_{i-1}\}$.  Then $a_i b_i$ is a nonedge and
$a_i b_j$ is an edge whenever $j<i$, so these vertices form a semi-ladder of
order $\ell$.
\end{proof}

The degree of a clause or constraint is its arity, while the degree of a ground
element is its frequency.  The degree of a variable is its variable frequency,
while the degree of a set is its size.  Throughout, $d$ denotes the stated
structural bound: a degree bound, the size of an excluded biclique, or a bound
on the semi-ladder index or VC dimension.  It is included in the parameter
$k+d$ and is not assumed to be an absolute constant unless stated otherwise.

For an instance $I$ and an explicit budget $q$, let $\OPT(I,q)$ be its minimum objective
value among solutions using at most $q$ selectable objects: sets in a coverage
instance or variables set to one in a SAT or CSP instance.  For a coverage instance
$I=(U,\mathcal S,w,k)$, we use the complementary notation
\[
  \OPT_{\mathrm{miss}}(I):=\OPT(I,k)
  \qquad\text{and}\qquad
  \OPT_{\mathrm{cov}}(I):=w(U)-\OPT_{\mathrm{miss}}(I).
\]
An \emph{efficient parameterized approximation scheme} (EPAS)
for one of the restricted problems takes $\eps>0$ and returns a feasible
solution of value at most $(1+\eps)\OPT(I,k)$.  Its running time is
\[
  f(k,d,1/\eps)N^{O(1)}
\]
for a computable function $f$, where $N$ denotes the input encoding size.
More generally, a \emph{finite-factor FPT
approximation} parameterized by $p$ runs in $g(p)N^{O(1)}$ time and returns a
feasible solution of value at most $c(p)\OPT(I,k)$, for some computable
functions $g$ and $c\colon\mathbb N\to[1,\infty)$.  Thus its approximation
factor may depend on the parameter but not on the input size.
All rational weights are encoded in binary and are
included in $N$.
Arithmetic and comparisons on these weights therefore take time polynomial in
the input size.

The following elementary proposition will be used in the hardness results.

\begin{proposition}[Zero-optimum barrier]
\label{prop:zero-barrier}
Let $\mathcal W\in\{\Wone,\Wtwo\}$.  Suppose it is $\mathcal W$-hard,
parameterized by $p$, to decide whether some feasible solution covers every
ground element or satisfies every constraint.  Unless $\FPT=\mathcal W$, the
corresponding minimization problem has no FPT $c(p)$-approximation
parameterized by $p$ for any computable function
$c\colon\mathbb N\to[1,\infty)$.
\end{proposition}

\begin{proof}
On a yes-instance the optimum is zero, so every solution satisfying a finite
multiplicative approximation guarantee must also have value zero.  On a
no-instance every feasible solution has positive value.  Running the
approximation algorithm and testing whether its output has value zero would
therefore decide the $\mathcal W$-hard problem in FPT time.
\end{proof}

All objective-value hardness proofs in this paper use this zero-optimum
argument.  They distinguish optimum zero from positive optimum.  These proofs
do not establish hardness under the promise that the optimum is positive, or
a multiplicative gap between two positive optimum values.

Combining \cref{prop:zero-barrier} with the standard $\Wtwo$-hardness of
\textsc{Set Cover} parameterized by the solution
size~\cite{cygan2015parameterized} yields the following folklore consequence.

\begin{corollary}[Folklore]
\label{cor:unrestricted-hardness}
Unless $\FPT=\Wtwo$, none of \MinUnCoverage,
\MinUnsatSAT, and \MinUnsatCSP{} admits a finite-factor FPT
approximation parameterized by $k$, even when every ground element, clause, or
constraint has weight one.
\end{corollary}

\begin{proof}
An instance of \textsc{Set Cover} has a cover of size at most $k$ if and only
if the corresponding \MinUnCoverage{} instance has optimum zero.  For
each ground element $u$, the monotone clause
$\bigvee_{S\in\mathcal S:u\in S}x_S$ is satisfied exactly when the selected
sets cover $u$.  Thus coverage is a special case of monotone SAT, and SAT is a
special case of Boolean CSP.
\end{proof}

The following is another folklore consequence of \cref{prop:zero-barrier},
now combined with the bounded-VC-dimension hardness of Bringmann et
al.~\cite{bringmann2016hittingset}.

\begin{corollary}[Folklore]
\label{thm:vc-dimension-hardness}
Unless $\FPT=\Wone$, none of \MinUnCoverage, \MinUnsatSAT, and
\MinUnsatCSP{} admits a finite-factor FPT approximation parameterized by
$k+d$ on incidence systems of VC dimension at most $d$, even when $d=2$ and
every ground element, clause, or constraint has weight one.
\end{corollary}

\begin{proof}
Bringmann et al.\ proved that \textsc{Hitting Set} is
$\Wone$-hard parameterized by the solution size even when both the primal and
dual VC dimensions are at most two~\cite{bringmann2016hittingset}.  Dualizing
such an instance gives a \textsc{Set Cover} instance of VC dimension at most
two with the same solution size.  By \cref{prop:zero-barrier}, this rules out a
finite-factor FPT approximation for \MinUnCoverage.  The monotone-CNF
encoding used in the proof of \cref{cor:unrestricted-hardness} has an
incidence graph isomorphic to that of the \textsc{Set Cover} instance: each
set $S$ becomes a variable $x_S$, each element $u$ becomes a clause $C_u$,
and $x_S$ occurs in $C_u$ exactly when $u\in S$.  Hence the VC dimension
remains at most two, and the same lower bound applies to \MinUnsatSAT{} and
therefore to \MinUnsatCSP.
\end{proof}

%% file: sections/bounded_arity_csp.tex
\section{An EPAS for Weighted \MinUnsatCSP{} with Bounded Arity}
\label{sec:bounded-arity}

Fix $d\geq1$ and $0<\eps\leq1$.  For a weighted Boolean CSP instance
$I=(X,\mathcal C,w)$ of arity at most $d$, we write $\zero$ for the all-zero
assignment on the current variable set and put
\[
  \mathcal Z(I)=\{C\in\mathcal C:\zero\text{ does not satisfy }C\},
  \qquad z_w(I)=\sum_{C\in\mathcal Z(I)}w(C).
\]
For $x\in X$, define its weighted occurrence in $\mathcal Z(I)$ by
\[
  \occ_{\mathcal Z(I),w}(x)
  =\sum_{\substack{C\in\mathcal Z(I)\\x\in\operatorname{scope}(C)}}w(C).
\]

\paragraph*{Algorithmic idea.}
The algorithm first considers the simplest feasible solution: assign zero to
every variable of the current instance.  This solution violates total
constraint weight $z_w(I)$.  The initial call has budget $q=k$; in every later
call, $q$ is the remaining number of variables that may still be set to one.  To
look for a better solution, the algorithm sets a variable $x$ to the Boolean
value one.  This uses one unit of the budget, so it solves the remaining
instance recursively with budget $q-1$.  Trying every variable would be too
expensive, so we use the constraints in $\mathcal Z(I)$ to select a small set
of variables.  Define
\[
  \theta(I,q)=\frac{\eps}{(1+\eps)q}z_w(I)
  \qquad\text{and}\qquad
  H(I,q)=\{x\in X:\occ_{\mathcal Z(I),w}(x)>\theta(I,q)\}.
\]
The algorithm tries setting each variable in $H(I,q)$ to one.  If the
all-zero solution is not a $(1+\eps)$-approximation, let $\mathcal R$ be the
constraints in $\mathcal Z(I)$ that an optimum assignment satisfies.  Their
total weight satisfies
$w(\mathcal R)\ge z_w(I)-\OPT(I,q)>\eps z_w(I)/(1+\eps)$.  Each constraint in
$\mathcal R$ contains a variable that the optimum sets to one.  Charge its
weight to one such variable.  Since the optimum sets at most $q$ variables to
one, some variable receives total charge at least
$w(\mathcal R)/q>\theta(I,q)$.  Its weighted occurrence is at least this
charge, so it belongs to
$H(I,q)$.  On the other hand, arity at most $d$ implies
$|H(I,q)|<d(1+\eps)q/\eps$.  Thus the algorithm tries only a bounded number of
variables, including an optimum choice whenever the all-zero solution is not
good enough, and returns the best solution found.  The residual construction
below defines the instance obtained after setting $x=1$, and
\cref{lem:csp-witness,lem:csp-candidates} prove the two stated properties of
$H(I,q)$.

\paragraph*{Residual instances.}
Fixing $x=1$ restricts every constraint containing $x$.  Formally, delete the
coordinate of $x$ from its scope.  If the relation is represented by its
satisfying tuples, retain the projected tuples whose extension by a one in
that coordinate satisfies the original relation.  If it is represented by
its falsifying tuples, apply the same operation to the falsifying list and
retain the complement representation.
Constraints with an empty scope are retained as constant true or constant
false constraints, and every residual constraint retains its weight.  Denote
the resulting instance by $I[x\gets1]$.  Every assignment $\sigma$ of the
residual variables satisfies
\begin{equation}
\label{eq:csp-residual}
  \unsat_I(\sigma\cup\{x\mapsto1\})
  =\unsat_{I[x\gets1]}(\sigma).
\end{equation}
This restriction operation does not increase arity and is polynomial in the
explicit input representation.

The following pseudocode implements this search.

\begin{algorithm}[H]
\caption{$\textsc{Min-Unsat-$d$-CSP}(I,q,\eps)$}
\label{alg:min-unsat-csp}
\begin{algorithmic}[1]
\Input A weighted Boolean CSP instance $I=(X,\mathcal C,w)$ of arity at most
$d$, a remaining budget $q$, and $\eps\in(0,1]$.
\Output An assignment of Hamming weight at most $q$.
\State $\tau_{\mathrm{best}}\gets\zero$
\If{$q=0$ or $z_w(I)=0$}
  \State \Return $\tau_{\mathrm{best}}$
\EndIf
\State $\theta\gets \eps z_w(I)/((1+\eps)q)$
\State $H\gets\{x\in X:\occ_{\mathcal Z(I),w}(x)>\theta\}$
\ForAll{$x\in H$}
  \State $\sigma_x\gets\textsc{Min-Unsat-$d$-CSP}(I[x\gets1],q-1,\eps)$
  \State $\tau_x\gets\sigma_x\cup\{x\mapsto1\}$
  \If{$\unsat_I(\tau_x)<\unsat_I(\tau_{\mathrm{best}})$}
    \State $\tau_{\mathrm{best}}\gets\tau_x$
  \EndIf
\EndFor
\State \Return $\tau_{\mathrm{best}}$
\end{algorithmic}
\end{algorithm}

We now prove that the branching set contains an optimum choice when branching
is necessary and that its size is bounded.

\begin{lemma}[An optimum branch exists]
\label{lem:csp-witness}
Let $I=(X,\mathcal C,w)$ be a weighted Boolean CSP instance of arity at most
$d$, let $q\ge1$, and let $\tau^*$ be an optimum assignment of Hamming weight
at most $q$.
If $z_w(I)>(1+\eps)\OPT(I,q)$, then
$\{x\in X:\tau^*(x)=1\}\cap H(I,q)\ne\emptyset$.
\end{lemma}

\begin{proof}
The constraints in $\mathcal Z(I)$ have total weight $z_w(I)$, and $\tau^*$
violates constraints of total weight at most $\OPT(I,q)$.  Therefore, the
constraints in $\mathcal Z(I)$ that it satisfies have total weight greater
than $\eps z_w(I)/(1+\eps)$.  For each of these satisfied constraints, choose a
variable in its scope that $\tau^*$ sets to one.  Such a variable must exist;
otherwise, $\tau^*$ and the all-zero assignment would assign the same tuple to
the constraint, and both would violate it.  Associate the weight of each such
constraint with one chosen variable.  Since $\tau^*$ sets at most $q$
variables to one, some variable receives total weight greater than
$\eps z_w(I)/((1+\eps)q)=\theta(I,q)$.  Its weighted occurrence is at least
this value.  Hence, this variable belongs to both
$\{x\in X:\tau^*(x)=1\}$ and
$H(I,q)$.
\end{proof}

\begin{lemma}[The branching set is small]
\label{lem:csp-candidates}
For every weighted Boolean CSP instance $I=(X,\mathcal C,w)$ of arity at most
$d$ and every $q\ge1$, $|H(I,q)|<d(1+\eps)q/\eps$.
\end{lemma}

\begin{proof}
If $z_w(I)=0$, then every weighted occurrence is zero, so $H(I,q)=\emptyset$
and the claim follows.  Assume henceforth that $z_w(I)>0$.
Each constraint in $\mathcal Z(I)$ contains at most $d$ variables.  Therefore,
$\sum_{x\in X}\occ_{\mathcal Z(I),w}(x)\le d\,z_w(I)$.  Moreover, every
$x\in H(I,q)$ satisfies $\occ_{\mathcal Z(I),w}(x)>\theta(I,q)
=\eps z_w(I)/((1+\eps)q)$.  Hence,
$|H(I,q)|\theta(I,q)<\sum_{x\in X}\occ_{\mathcal Z(I),w}(x)
\le d\,z_w(I)$.
Substituting the value of $\theta(I,q)$ and rearranging gives
$|H(I,q)|<d(1+\eps)q/\eps$, as claimed.
\end{proof}

\begin{lemma}[Approximation guarantee]
\label{lem:csp-approximation}
For every weighted Boolean CSP instance $I$ of arity at most $d$ and every
$q\ge0$, \cref{alg:min-unsat-csp} returns an assignment $\tau$ of Hamming
weight at most $q$ with
$\unsat_I(\tau)\le(1+\eps)\OPT(I,q)$.
\end{lemma}

\begin{proof}
We prove the claim simultaneously for every weighted arity-$d$ instance by
induction on $q$.  For the base case $q=0$, the all-zero assignment is the only
feasible assignment, and the algorithm returns it.  Now fix $q\ge1$ and assume
as the induction hypothesis that the claim holds for every weighted
arity-$d$ instance with budget $q-1$.  If $z_w(I)=0$, the all-zero assignment
has objective value zero and is therefore optimal, and the algorithm returns
it.  We may therefore assume that $z_w(I)>0$.

First suppose that $z_w(I)\le(1+\eps)\OPT(I,q)$.  The all-zero assignment
violates total constraint weight exactly $z_w(I)$, so it satisfies the required
approximation guarantee.  Since the algorithm returns the best assignment it
finds, its output also satisfies this guarantee.

Now suppose that $z_w(I)>(1+\eps)\OPT(I,q)$.  By \cref{lem:csp-witness}, there
is a variable $x^*\in H(I,q)$ that an optimum assignment $\tau^*$ sets to one.
The algorithm therefore considers the branch $x^*=1$.  Restricting $\tau^*$
to the remaining variables gives an assignment of Hamming weight at most
$q-1$ for $I[x^*\gets1]$ with value $\OPT(I,q)$.  Conversely, any better
assignment for the residual instance would extend to an assignment for $I$
with Hamming weight at most $q$, contradicting the optimality of $\tau^*$.
Thus, the optimum value
of the residual instance is $\OPT(I,q)$.

By the induction hypothesis, the recursive call returns an assignment of
value at most $(1+\eps)\OPT(I,q)$.  By \cref{eq:csp-residual}, extending this
assignment with $x^*=1$ gives the same value for $I$.  Since the algorithm
returns the best candidate, the claim follows.
\end{proof}

\begin{theorem}
\label{thm:arity-csp}
For every $d\ge1$, weighted arity-$d$ \MinUnsatCSP{} admits a deterministic
EPAS parameterized by $k+d$.  For $0<\eps\le1$, its running time is
$(dk/\eps)^{O(k)}N^{O(1)}$.
\end{theorem}

\begin{proof}
It is enough to consider $\eps\in(0,1]$.  If $\eps>1$, we run the algorithm
with $\eps'=1$; the resulting approximation factor of $2$ is at most
$1+\eps$.  We start the algorithm with budget $q=k$, and
\cref{lem:csp-approximation} gives the required approximation guarantee.

Each recursive call reduces the remaining budget by one.  Since the initial
budget is $k$, the recursion has at most $k$ levels.  Let
$c=\lceil d(1+\eps)/\eps\rceil$, and let $N_i$ be the number of recursive
calls at depth $i$, where the initial call has depth zero.  By
\cref{lem:csp-candidates}, a call with remaining budget $q$ considers fewer
than $d(1+\eps)q/\eps\le cq$ variables.  Therefore, it makes at most $cq$
recursive calls.

We claim that $N_i\le c^i k!/(k-i)!$ for every $i\in\{0,\ldots,k\}$.  At
depth zero, there is only the initial call, so $N_0=1$, as required.  Now
suppose the claim holds at depth $i$.  Every call at this depth has remaining
budget $k-i$ and therefore makes at most $c(k-i)$ recursive calls.  Hence,
\[
\begin{aligned}
  N_{i+1}
  &\le N_i\,c(k-i)
  \le c^i\frac{k!}{(k-i)!}\,c(k-i)
  =c^{i+1}\frac{k!}{(k-i-1)!}.
\end{aligned}
\]
This proves the claim by induction.

The total number of calls is therefore at most
\[
\begin{aligned}
  \sum_{i=0}^{k}N_i
  &\le \sum_{i=0}^{k}c^i\frac{k!}{(k-i)!}
  =c^k k!\sum_{j=0}^{k}\frac{1}{c^j j!}.
\end{aligned}
\]
Since $c\ge2$ and $j!\ge1$, we have
$\sum_{j=0}^{k}1/(c^j j!)\le\sum_{j=0}^{k}1/2^j<2$.  Thus, the total number
of recursive calls is less than $2c^k k!$, and hence it is $O(c^k k!)$.

Constructing a residual instance and comparing candidate assignments takes
polynomial time.  Since $c=O(d/\eps)$ and $k!\le k^k$, we have
$c^k k!\le(dk/\eps)^{O(k)}$.  Therefore, the total running time is
$(dk/\eps)^{O(k)}N^{O(1)}$, as stated.
\end{proof}

\begin{corollary}
\label{cor:arity-special-cases}
Weighted arity-$d$ \MinUnsatSAT{} and weighted \MinUnCoverage{} with
ground-element frequency at most $d$ admit deterministic EPAS algorithms
parameterized by $k+d$.
\end{corollary}

\begin{proof}
By \cref{thm:arity-csp}, weighted bounded-arity \MinUnsatCSP{} admits a
deterministic EPAS.  Since CNF formulas are a special case of Boolean CSP, the result for
arity-$d$ \MinUnsatSAT{} follows immediately.

We obtain the coverage result by reducing \MinUnCoverage{} to
\MinUnsatSAT{}.  Given an instance $(U,\mathcal S,w,k)$, introduce a Boolean
variable $x_S$ for each set $S\in\mathcal S$, where $x_S=1$ means that $S$ is
selected.  For each ground element $u$, add the monotone clause
$\bigvee_{S\in\mathcal S:\,u\in S}x_S$ with weight $w(u)$.  An assignment of
Hamming weight at most $k$ corresponds to a subfamily of at most $k$ sets.  The
clause associated with
$u$ is violated exactly when $u$ is uncovered, so the reduction preserves the
objective value.  If $u$ belongs to no input set, this disjunction is empty and
denotes the always-false clause.  Its arity equals the number of sets containing
$u$, so the reduction preserves the bound $d$.  The coverage result follows from
\cref{thm:arity-csp}.

Skowron and Faliszewski previously obtained the bounded-frequency coverage
result for unit weights using a randomized
algorithm~\cite{skowron2017boundedfrequencies}; the algorithm obtained here is
deterministic and allows rational objective weights.
\end{proof}

\begin{proposition}[Exactly $k$ choices]
\label{prop:exact-cardinality}
For \MinUnCoverage{}, the at-most-$k$ and exactly-$k$ variants have the same
optimum whenever $|\mathcal S|\geq k$, and all objective-value coverage
algorithms in this paper extend to the exactly-$k$ variant.  If
$|\mathcal S|<k$, that variant is infeasible.
In contrast, unless $\FPT=\Wone$, neither \MinUnsatSAT{} nor
\MinUnsatCSP{} with exactly $k$ true variables admits a finite-factor FPT
approximation parameterized by $k$, even for unweighted clauses of arity two.
\end{proposition}

\begin{proof}
For coverage, a subfamily of fewer than $k$ sets can be extended to $k$ sets
without increasing the uncovered weight.  This proves equality of the two
optima and preserves the guarantee of any approximation algorithm.

For SAT, take an instance $(G,k)$ of \textsc{Independent Set}, which is
$\Wone$-hard parameterized by $k$~\cite{cygan2015parameterized}.  We may assume
$k\leq |V(G)|$.  Introduce a variable $x_v$ for each vertex and the unit-weight
clause $(\neg x_u\vee\neg x_v)$ for each edge $uv$.  An assignment with
exactly $k$ true variables satisfies every clause if and only if those
variables represent an independent set of size $k$.  Thus its minimum
unsatisfied weight is zero precisely on yes-instances.  By the zero-optimum
argument in \cref{prop:zero-barrier}, a finite-factor FPT approximation would
decide \textsc{Independent Set} in FPT time.  The CSP claim follows because
SAT is a special case of Boolean CSP.
\end{proof}

%% file: sections/kdd_hardness.tex
\section{Hardness for \MinUnsatSAT{} and \MinUnsatCSP{}}
\label{sec:kdd-hardness}

The previous section gives an EPAS for weighted instances when every constraint
has bounded arity. In this section, we show that this positive result does not
extend to general CNF formulas under biclique-freeness, bounded semi-ladder
index, or bounded variable frequency.

\subsection{Hardness under Biclique-Freeness and Bounded Semi-Ladder Index}
We give a parameter-preserving reduction from
\textsc{Multicolored Independent Set}, which is $\Wone$-hard when
parameterized by the number of colors~\cite{cygan2015parameterized} as follows.

\paragraph*{Construction.}
Let $(G,V_1,\ldots,V_k)$ be an instance of \textsc{Multicolored Independent Set}, where
$V(G)=V_1\uplus\cdots\uplus V_k$.  Introduce a Boolean variable $x_v$ for each
vertex $v\in V(G)$.  We may delete all edges within a color class, since they
are irrelevant to a multicolored independent set.  For each color $i\in[k]$,
add the selection clause $A_i=\bigvee_{v\in V_i}x_v$.  For each edge
$uv\in E(G)$, add the conflict clause
$B_{uv}=\neg x_u\vee\neg x_v$.  Finally, for each $i\in[k]$ and each pair of
distinct vertices $u,v\in V_i$, add the pairwise exclusion clause
$D_{uv}=\neg x_u\vee\neg x_v$.  Let $\Phi_G$ be the conjunction of these
clauses, and set the Hamming-weight bound to $k$.
\begin{claim}
\label{clm:k32-incidence}
The incidence graph of $\Phi_G$ excludes both orientations of $K_{3,2}$ and,
in particular, is $K_{3,3}$-free.
\end{claim}

\begin{proof}
Every clause other than a selection clause contains only two variables, and
all variables in a selection clause belong to the same color class.  Thus any
three variables have at most one common clause neighbor, namely their common
selection clause when all three belong to the same color class. Hence the incidence graph contains no
$K_{3,2}$ whose three-vertex side lies in the variable part.

Conversely, any two variables have at most two common clause neighbors. Two
variables from the same color class occur together only in their selection
clause and their pairwise exclusion clause, while two variables from different
color classes occur together in at most one conflict clause. Hence the
incidence graph contains no $K_{2,3}$ whose three-vertex side lies in the
clause part.
Thus both orientations of $K_{3,2}$ are excluded. In particular, the
incidence graph is $K_{3,3}$-free.
\end{proof}
\begin{claim}
\label{clm:kdd-sat-equivalence}
$G$ contains a multicolored independent set of size $k$ if and only if
$\Phi_G$ has a satisfying assignment.  Moreover, every satisfying assignment
of $\Phi_G$ has Hamming weight exactly $k$.
\end{claim}
\begin{proof}
($\Rightarrow$) Suppose $S$
is a multicolored independent set in $G$.  Set $x_v=1$ exactly when $v\in S$.
This assignment has Hamming weight $k$ and satisfies
every selection clause and every pairwise exclusion clause.  It also satisfies
every conflict clause because no edge has both endpoints in $S$.  Hence, it
satisfies $\Phi_G$.

($\Leftarrow$) Conversely, let $\tau$ be a satisfying assignment.  Each selection clause
$A_i$ forces at least one variable from $V_i$ to be true, and the pairwise
exclusion clauses force at most one such variable to be true.  Hence every
satisfying assignment, even without a Hamming-weight restriction, sets exactly
one variable from each color class to one and has Hamming weight exactly $k$.
Let $S$ be the corresponding set of vertices.  Every conflict clause is
satisfied, so no edge has both endpoints in $S$.  Therefore, $S$ is a
multicolored independent set.

\end{proof}

\begin{theorem}
\label{thm:kdd-sat-hard}
Unless $\FPT=\Wone$, for every computable function
$g\colon\mathbb N^2\to\mathbb R_{\geq 1}$, neither \MinUnsatSAT{} nor
\MinUnsatCSP{} admits a $g(k,d)$-approximation in FPT time parameterized by
$k+d$ on unweighted $K_{d,d}$-free instances or on unweighted instances of
semi-ladder index at most $d$. In particular, the hardness holds already on
$K_{3,3}$-free instances and on instances of semi-ladder index at most four,
respectively, and hence neither problem admits a finite-factor FPT
approximation parameterized by $k$ on these classes.

Moreover, assuming ETH, neither problem admits a finite-factor approximation
running in $f(k,d)N^{o(k+d)}$ time under either restriction, for any
computable function $f$.
\end{theorem}

\begin{proof}
Give every clause unit weight.  By
\cref{clm:k32-incidence,clm:kdd-sat-equivalence}, the construction is a
parameter-preserving reduction from \textsc{Multicolored Independent Set},
the resulting incidence graph excludes both orientations of $K_{3,2}$, and
every satisfying assignment has Hamming weight exactly $k$.
Thus \cref{prop:zero-barrier} gives the claimed approximation hardness for
$K_{3,3}$-free \MinUnsatSAT{}.

A bipartite graph excluding both orientations of $K_{3,2}$ has semi-ladder
index at most four: a semi-ladder of order five contains all edges between
$\{a_3,a_4,a_5\}$ and $\{b_1,b_2\}$, and hence contains a $K_{3,2}$.
This proves the semi-ladder claim.  The same lower bounds hold for
\MinUnsatCSP{} because \MinUnsatSAT{} is a special case of Boolean CSP that
preserves the incidence graph.

For the ETH lower bound, \textsc{Independent Set} on an $n$-vertex graph has
no $f(k)n^{o(k)}$-time algorithm for any computable function
$f$~\cite{chen2006strong}; the standard construction with $k$ colored copies
gives the same lower bound for \textsc{Multicolored Independent Set}.  Our
reduction produces a formula of size $N=n^{O(1)}$ and preserves $k$.  Since
$d=3$ in the biclique-free case and $d=4$ in the semi-ladder case,
$k+d=\Theta(k)$, which yields the stated running-time lower bound.
\end{proof}

\subsection{Hardness under Bounded Variable Frequency}
We next show that the positive result for bounded arity does not extend to
bounded variable frequency. We give a parameter-preserving reduction from
\textsc{Multicolored Clique}, which is $\Wone$-hard when parameterized by the
number of colors~\cite{cygan2015parameterized}.

\paragraph*{Construction.}
Let $(G,V_1,\ldots,V_r)$ be an instance of
\textsc{Multicolored Clique}, where
$V(G)=V_1\uplus\cdots\uplus V_r$. Introduce a variable $x_v$ for every
vertex $v$ and a variable $y_e$ for every edge $e$ joining two different
color classes.

\begin{itemize}
    \item For every $i\in[r]$, add the vertex-selection clause
    \[
    X_i=\bigvee_{v\in V_i}x_v.
    \]

    \item For every $1\le i<j\le r$, add the edge-selection clause
    \[
    Y_{ij}=\bigvee_{e\in E(V_i,V_j)}y_e.
    \]
    If some $E(V_i,V_j)$ is empty, then the source instance is a
    no-instance; in this case, we output the single clause $(z)$ with
    Hamming-weight bound zero. We may therefore assume that every
    edge-selection clause is nonempty.

    \item For every $v\in V_i$ and every $j\ne i$, add the consistency
    clause
    \[
    C(v,j)=\neg x_v\vee
    \bigvee_{e=vw,\,w\in V_j}y_e.
    \]
    If $v$ has no neighbor in $V_j$, this clause is simply $\neg x_v$.

    \item Set the Hamming-weight bound to
    \[
    k':=r+\binom r2,
    \]
    and let $\Phi_G$ be the conjunction of all the clauses above.
\end{itemize}

\begin{claim}
\label{clm:frequency-bound}
The maximum variable frequency of $\Phi_G$ is at most
$d:=\max\{r,3\}$, and $k'+d=O(r^2)$.
\end{claim}

\begin{proof}
Each variable $x_v$, where $v\in V_i$, occurs in $X_i$ and in the $r-1$
clauses $C(v,j)$ with $j\ne i$, and hence has frequency $r$. If $e=uv$
joins $V_i$ and $V_j$, then $y_e$ occurs in exactly three clauses:
$Y_{ij}$, $C(u,j)$, and $C(v,i)$. Thus the maximum variable frequency is at
most $d:=\max\{r,3\}$. Since
$k'=r+\binom r2=O(r^2)$, we also have $k'+d=O(r^2)$.
\end{proof}

\begin{claim}
\label{clm:frequency-equivalence}
$G$ contains a multicolored clique of size $r$ if and only if $\Phi_G$ has
a satisfying assignment of Hamming weight at most $k'$.
\end{claim}

\begin{proof}
($\Rightarrow$) Suppose that
$\{v_1,\ldots,v_r\}$ is a multicolored clique, where $v_i\in V_i$.
Set $x_{v_i}=1$ for every $i\in[r]$ and
$y_{v_iv_j}=1$ for every $1\le i<j\le r$, and set all other variables to
zero. This assignment has Hamming weight exactly $k'$. Every
vertex-selection clause $X_i$ and every edge-selection clause $Y_{ij}$ is
satisfied. Moreover, for every selected vertex $v_i$ and every $j\ne i$,
the variable $y_{v_iv_j}$ occurs in $C(v_i,j)$ and is set to one, while
every consistency clause corresponding to an unselected vertex is satisfied
by its negative literal. Hence $\Phi_G$ is satisfied.

($\Leftarrow$) Conversely, suppose that $\Phi_G$ has a satisfying assignment
of Hamming weight at most $k'$. There are exactly
$r+\binom r2=k'$ selection clauses, namely the clauses $X_i$ and $Y_{ij}$,
and their variable sets are pairwise disjoint. Hence satisfying all of them
requires at least $k'$ true variables. The Hamming-weight bound therefore
implies that exactly one variable is set to one in each selection clause and
no other variable is set to one.

For every $i\in[r]$, let $x_{v_i}$ be the unique true variable in $X_i$,
and for every $1\le i<j\le r$, let $y_{e_{ij}}$ be the unique true variable
in $Y_{ij}$. Since $x_{v_i}=1$, the clause $C(v_i,j)$ can be satisfied only
by a true edge variable representing an edge from $v_i$ to $V_j$. Since
$y_{e_{ij}}$ is the unique true variable in $Y_{ij}$, the edge $e_{ij}$ must
be incident with $v_i$. Similarly, $C(v_j,i)$ implies that $e_{ij}$ is
incident with $v_j$. Hence $e_{ij}=v_iv_j$. This holds for every pair
$i<j$, and therefore $\{v_1,\ldots,v_r\}$ is a multicolored clique.
\end{proof}

\begin{theorem}
\label{thm:frequency-sat-hard}
Unless $\FPT=\Wone$, for every computable function
$g\colon\mathbb N^2\to\mathbb R_{\geq 1}$, neither \MinUnsatSAT{} nor
\MinUnsatCSP{} admits a $g(k,d)$-approximation in FPT time parameterized by
$k+d$ on unweighted instances of maximum variable frequency at most $d$.

Moreover, assuming ETH, neither problem admits a finite-factor approximation
running in $f(k,d)N^{o(\sqrt{k+d})}$ time for any computable function $f$.
\end{theorem}

\begin{proof}
By \cref{clm:frequency-bound,clm:frequency-equivalence}, the construction is
a parameterized reduction from \textsc{Multicolored Clique}: the resulting
formula is satisfiable by an assignment of Hamming weight at most $k'$ if and
only if the source graph has a multicolored clique, and its maximum variable
frequency is at most $d=\max\{r,3\}$.  Giving every clause unit weight and
applying \cref{prop:zero-barrier} proves the approximation hardness for
\MinUnsatSAT{}.  It extends to \MinUnsatCSP{} because SAT is a special case of
Boolean CSP and the encoding preserves variable frequency.

For the ETH lower bound, \textsc{Clique} on an $n$-vertex graph has no
$f(r)n^{o(r)}$-time algorithm for any computable function
$f$~\cite{chen2006strong}; the standard construction with $r$ colored copies
gives the same lower bound for \textsc{Multicolored Clique}.  Our reduction
produces a formula of size $N=n^{O(1)}$ with
$k'+d=\Theta(r^2)$.  Therefore, a running time of
$f(k,d)N^{o(\sqrt{k+d})}$ would yield an $f'(r)n^{o(r)}$-time algorithm for
\textsc{Multicolored Clique}, contradicting ETH.
\end{proof}

%% file: sections/semiladder_uncoverage.tex
\section[An EPAS for Weighted \MinUnCoverage{} with Bounded Semi-Ladder Index ]
{An EPAS for Weighted \MinUnCoverage{} with Bounded Semi-Ladder Index }
\label{sec:semiladder-uncoverage}

We now move beyond bounded ground-element frequency and place a restriction
on the whole incidence graph.  Fix $\eps\in(0,1]$ and consider a weighted
\MinUnCoverage{} instance $I=(U,\mathcal S,w,k)$ whose incidence graph has
semi-ladder index at most $d$.  Write
\[
  \OPT_{\mathrm{miss}}(I)
  =\min_{\substack{\mathcal A\subseteq\mathcal S\\|\mathcal A|\le k}}
    \miss_{\mathcal S,w}(\mathcal A).
\]

We adapt Kundu et al.'s skeleton-based branching framework and deterministic
weighted-net construction~\cite{kundu2026fixedbudget}.  The skeletons,
eligible families, forced-coverage intersections, weighted nets, and
semi-ladder depth argument come from that framework.  Our contribution here
is the residual-weight argument: unless the current candidate has the required
approximation guarantee, an optimum set covers enough unresolved weight to
be hit by the net.  This yields an approximation to the uncovered weight
while keeping the original budget of at most $k$ sets.

Assume for now that $k\ge1$ and $\mathcal S\ne\emptyset$; the remaining cases
are immediate.  The algorithm maintains an ordered tuple of skeletons
\[
  \mathbf M=(M_1,\ldots,M_k), \qquad M_i\subseteq U.
\]
The skeleton $M_i$ records the elements that the set chosen for coordinate $i$
must contain.  Its eligible choices are
\[
  \mathcal P_i(\mathbf M)
  =\{S\in\mathcal S:M_i\subseteq S\}
\]
and the state is feasible when every $\mathcal P_i(\mathbf M)$ is nonempty.
At a feasible state, define
\[
  F_i(\mathbf M)=\bigcap_{S\in\mathcal P_i(\mathbf M)}S,
  \qquad
  F(\mathbf M)=\bigcup_{i=1}^k F_i(\mathbf M),
  \qquad
  W(\mathbf M)=U\setminus F(\mathbf M).
\]
Here $F_i(\mathbf M)$ is the coverage forced at coordinate $i$,
$F(\mathbf M)$ is the coverage guaranteed by all coordinates, and
$W(\mathbf M)$ contains the elements whose coverage remains unresolved.
Choosing any $A_i\in\mathcal P_i(\mathbf M)$ for every $i\in[k]$ gives a
candidate subfamily $\mathcal A=\{A_1,\ldots,A_k\}$.  It covers
$F(\mathbf M)$ and therefore leaves uncovered weight at most
$w(W(\mathbf M))$.

Because adding a set cannot increase the uncovered weight, an optimum
solution can be represented by an ordered $k$-tuple
$\mathbf O=(O_1,\ldots,O_k)$, allowing repeated sets.  We call a state
$\mathbf M$ \emph{consistent} with $\mathbf O$ if $M_i\subseteq O_i$, or
equivalently $O_i\in\mathcal P_i(\mathbf M)$, for every $i\in[k]$.

Branching depends only on how eligible sets meet the unresolved elements.  We
record these intersections in the restricted set system
\[
  \mathcal R(\mathbf M)
  =\left(W(\mathbf M),
    \left\{S\cap W(\mathbf M):
      S\in\bigcup_{i=1}^k\mathcal P_i(\mathbf M)\right\}\right).
\]
Intersecting the eligible sets with $W(\mathbf M)$ cannot increase the
semi-ladder index.  Suppose that distinct members $T_1,\ldots,T_\ell$ of the
restricted family and elements
$u_1,\ldots,u_\ell\in W(\mathbf M)$ form a semi-ladder in
$\mathcal R(\mathbf M)$.  For each $i$, choose an eligible set $S_i$ such that
$T_i=S_i\cap W(\mathbf M)$.  The sets $S_1,\ldots,S_\ell$ are distinct because
the sets $T_1,\ldots,T_\ell$ are distinct.  Since every $u_j$ belongs to
$W(\mathbf M)$, we
have $u_j\in T_i$ if and only if $u_j\in S_i$.  Hence
$u_i\notin S_i$ and $u_j\in S_i$ whenever $j<i$.  Thus,
$S_1,\ldots,S_\ell$ and $u_1,\ldots,u_\ell$ form a semi-ladder of order
$\ell$ in the original incidence graph.  Therefore,
$\mathcal R(\mathbf M)$ has semi-ladder index at most $d$.

By \cref{obs:vc-semiladder}, semi-ladder index at most $d$ implies VC
dimension at most $d$.  Thus standard weighted $\delta$-net bounds already
explain the existence and size of the net used below.  The additional
algorithmic ingredient needed here is a deterministic FPT-time construction.
Kundu et al.\ obtain such a construction under bounded semi-ladder index using
Guillemot's FPT algorithm for \textsc{Set Cover}
\cite{kundu2026fixedbudget,guillemot2025parameterized}.
For $W\ne\emptyset$ and $0<\delta\le1$, a weighted
$\delta$-net for $(W,\mathcal P,w)$ is a set $Z\subseteq W$ that intersects
every $P\in\mathcal P$ with $w(P)\ge\delta w(W)$.  The following lemma is a
direct consequence of Kundu et al.'s Theorem~5.9 and Lemmas~8.10 and~8.12
\cite{kundu2026fixedbudget}, applied to
$\mathcal P_\delta=\{P\in\mathcal P:w(P)\ge\delta w(W)\}$.  Restricting
$\mathcal P$ to this subfamily cannot increase its semi-ladder index.

\begin{lemma}[Deterministic weighted nets for bounded semi-ladder index]
\label{lem:semiladder-net}
Let $(W,\mathcal P)$ be a set system with $W\ne\emptyset$ whose incidence
graph has semi-ladder index at most $d$, let $w\colon W\to\mathbb Q_{>0}$, and let
$\delta\in(0,1/2]$.  A weighted $\delta$-net of size
\[
  O\left(\frac{d+1}{\delta}\log\frac1\delta\right)
\]
can be found in time
\[
  \left(\frac{d+1}{\delta}\right)^{
    O\left(\frac{(d+1)^2}{\delta}\log\frac1\delta\right)
  }N^{O(1)},
\]
where $N$ is the encoding size of $(W,\mathcal P,w)$.
\end{lemma}

The polynomial factor accounts for exact arithmetic on binary rational
weights.  To form $\mathcal P_\delta$, we sum the weights in each range and
compare the result with $\delta w(W)$.  Each such sum has bit length
polynomial in the input size.  The remaining net construction is an
unweighted covering computation on the incidence graph of
$(W,\mathcal P_\delta)$.

Now consider a state consistent with $\mathbf O$ such that
$\sum_{i=1}^k|M_i|<kd$.  Since $O_i\in\mathcal P_i(\mathbf M)$, we have
$F_i(\mathbf M)\subseteq O_i$.  Hence every element
missed by $\mathbf O$ belongs to $W(\mathbf M)$, and these elements have total
weight $\OPT_{\mathrm{miss}}(I)$.  If
$w(W(\mathbf M))\le(1+\eps)\OPT_{\mathrm{miss}}(I)$, the current candidate
already has the required value.  Otherwise,
$w(W(\mathbf M))>(1+\eps)\OPT_{\mathrm{miss}}(I)$, so
$\OPT_{\mathrm{miss}}(I)<w(W(\mathbf M))/(1+\eps)$.  Hence the optimum covers
more than
$\eps w(W(\mathbf M))/(1+\eps)$ weight among the unresolved elements.
Therefore, one of its $k$ sets satisfies
\[
  w(O_i\cap W(\mathbf M))
  >\frac{\eps}{k(1+\eps)}w(W(\mathbf M)).
\]
Set $\delta=\eps/(k(1+\eps))$.  Since $k\ge1$ and $0<\eps\le1$, we have
$0<\delta\le1/2$.  By \cref{lem:semiladder-net}, a weighted $\delta$-net for
$\mathcal R(\mathbf M)$ contains an element
$u\in O_i\cap W(\mathbf M)$.  The algorithm does not know the optimum tuple,
so it tries every pair $(i,u)$ and restricts coordinate $i$ to eligible sets
that contain $u$.  The branch indexed by this pair $(i,u)$ remains consistent.

The comparison with $\OPT_{\mathrm{miss}}(I)$ is used only in the proof.
The algorithm does not know this value.  It records the candidate at every
state, explores the branches described below, and returns the best candidate
found.

Adding $u$ to $M_i$ also creates a new semi-ladder rung.  Indeed,
$u\in W(\mathbf M)$ implies $u\notin F_i(\mathbf M)$, so some eligible set
misses $u$ while containing every element already in $M_i$.  Each skeleton
therefore grows at most $d$ times, and the recursion depth is at most $kd$.
The complete procedure is given in \cref{alg:semiladder-uncoverage}.
\Cref{lem:semiladder-skeleton-depth,lem:semiladder-uncoverage-approximation}
establish the skeleton-size and approximation bounds.

\begin{algorithm}[t]
\caption{$\textsc{SL-MinUnCov}(I,\mathbf M,d,\eps)$}
\label{alg:semiladder-uncoverage}
\begin{algorithmic}[1]
\Input A weighted \MinUnCoverage{} instance $I=(U,\mathcal S,w,k)$ of
semi-ladder index at most $d$, skeletons
$\mathbf M=(M_1,\ldots,M_k)$, and $\eps\in(0,1]$.
\Output A subfamily of at most $k$ sets, or $\bot$ if some skeleton has no
eligible set.
\For{$i\gets1$ to $k$}
  \State $\mathcal P_i\gets\{S\in\mathcal S:M_i\subseteq S\}$
  \If{$\mathcal P_i=\emptyset$}
    \State \Return $\bot$
  \EndIf
  \State Choose any $A_i\in\mathcal P_i$
  \State $F_i\gets\bigcap_{S\in\mathcal P_i}S$
\EndFor
\State $\mathcal A\gets\{A_1,\ldots,A_k\}$,
$F\gets\bigcup_{i=1}^kF_i$, and $W\gets U\setminus F$
\If{$W=\emptyset$ or $\sum_{i=1}^k|M_i|\ge kd$}
  \State \Return $\mathcal A$
\EndIf
\State $\delta\gets\eps/(k(1+\eps))$
\State Use \cref{lem:semiladder-net} to compute a weighted $\delta$-net
$Z\subseteq W$ for
$\{S\cap W:S\in\bigcup_{i=1}^k\mathcal P_i\}$
\State $\mathcal A_{\mathrm{best}}\gets\mathcal A$
\ForAll{$(i,u)\in[k]\times Z$}
  \If{$\{S\in\mathcal P_i:u\in S\}\ne\emptyset$}
    \State Let $\mathbf M'$ be obtained from $\mathbf M$ by replacing
    $M_i$ with $M_i\cup\{u\}$
    \State $\mathcal B\gets\textsc{SL-MinUnCov}(I,\mathbf M',d,\eps)$
    \If{$\mathcal B\ne\bot$ and
    $\miss_{\mathcal S,w}(\mathcal B)
      <\miss_{\mathcal S,w}(\mathcal A_{\mathrm{best}})$}
      \State $\mathcal A_{\mathrm{best}}\gets\mathcal B$
    \EndIf
  \EndIf
\EndFor
\State \Return $\mathcal A_{\mathrm{best}}$
\end{algorithmic}
\end{algorithm}

The algorithm is invoked as
$\textsc{SL-MinUnCov}(I,(\emptyset,\ldots,\emptyset),d,\eps)$.  The test
before each recursive call ensures that the updated skeleton has an eligible
set; the eligible sets at the other coordinates do not change.

We first prove the depth bound.

\begin{lemma}[Skeleton size]
\label{lem:semiladder-skeleton-depth}
Consider any recursive call reached by
\cref{alg:semiladder-uncoverage}, and fix $h\in[k]$.  Write
$M_h=\{u_1,\ldots,u_\ell\}$ in the order in which its elements were inserted.
Then there exist pairwise distinct sets $R_1,\ldots,R_\ell\in\mathcal S$ such
that, for every $r\in[\ell]$,
\[
  u_r\notin R_r
  \qquad\text{and}\qquad
  u_s\in R_r\quad\text{for every }s<r.
\]
Thus, $R_1,\ldots,R_\ell$ and $u_1,\ldots,u_\ell$ form a semi-ladder of order
$\ell$.  Consequently, $|M_h|\le d$ for every $h\in[k]$.
\end{lemma}

\begin{proof}
Fix a coordinate $h$.  Suppose that the successive elements inserted into
$M_h$ are $u_1,\ldots,u_\ell$.  For each $r\in[\ell]$, let
$\mathbf M^{(r-1)}$ be the state immediately before $u_r$ is inserted into
coordinate $h$, and let $M_h^{(r-1)}$ be its $h$th skeleton.  The algorithm
chooses $u_r\in W(\mathbf M^{(r-1)})$, and hence
\[
  u_r\notin F_h(\mathbf M^{(r-1)})
  =\bigcap_{S\in\mathcal P_h(\mathbf M^{(r-1)})}S.
\]
Therefore, some set
$R_r\in\mathcal P_h(\mathbf M^{(r-1)})$ misses $u_r$.  At that time,
$\{u_1,\ldots,u_{r-1}\}\subseteq M_h^{(r-1)}$.  Every set in
$\mathcal P_h(\mathbf M^{(r-1)})$ contains $M_h^{(r-1)}$, so
\[
  u_r\notin R_r,
  \qquad
  u_s\in R_r \quad\text{whenever $s<r$}.
\]
Moreover, if $s<r$, then
$u_s\in M_h^{(r-1)}\subseteq F_h(\mathbf M^{(r-1)})$, whereas
$u_r\in W(\mathbf M^{(r-1)})\subseteq
U\setminus F_h(\mathbf M^{(r-1)})$.  Thus, $u_1,\ldots,u_\ell$ are pairwise
distinct.  The witness sets are also pairwise distinct: if $s<r$, then
$u_s\in R_r$ but $u_s\notin R_s$, and hence $R_r\ne R_s$.  Therefore,
$R_1,\ldots,R_\ell$ and $u_1,\ldots,u_\ell$ form a semi-ladder of order
$\ell$.  Since the incidence graph has semi-ladder index at most $d$, we have
$\ell\le d$.
\end{proof}

\begin{lemma}[Approximation guarantee]
\label{lem:semiladder-uncoverage-approximation}
Let $d\ge0$ and $0<\eps\le1$, and let
$I=(U,\mathcal S,w,k)$ be a weighted \MinUnCoverage{} instance with
$k\ge1$ and $\mathcal S\ne\emptyset$.  If its incidence graph has semi-ladder
index at most $d$, then
\[
  \textnormal{\textsc{SL-MinUnCov}}
  (I,(\emptyset,\ldots,\emptyset),d,\eps)
\]
returns a subfamily $\mathcal A\subseteq\mathcal S$ satisfying
\[
  |\mathcal A|\le k
  \qquad\text{and}\qquad
  \miss_{\mathcal S,w}(\mathcal A)
  \le(1+\eps)\OPT_{\mathrm{miss}}(I).
\]
\end{lemma}

\begin{proof}
List the sets of an optimum subfamily and append arbitrary sets from
$\mathcal S$ until the list has length $k$,
allowing repetitions.  This is possible because $\mathcal S\ne\emptyset$.
Adding sets cannot increase the uncovered weight, so the resulting tuple
$\mathbf O=(O_1,\ldots,O_k)$ remains optimal.
A state $\mathbf M$ is consistent with $\mathbf O$ if
$O_i\in\mathcal P_i(\mathbf M)$ for every $i\in[k]$.  In particular, the
initial state $(\emptyset,\ldots,\emptyset)$ is consistent.
A state is \emph{reachable} if it occurs in a recursive call generated from
this initial state.

For a reachable state $\mathbf M$, define
\[
  q(\mathbf M):=kd-\sum_{h=1}^k|M_h|.
\]
By \cref{lem:semiladder-skeleton-depth}, $q(\mathbf M)\ge0$.  For
$r\in\{0,\ldots,kd\}$, let $P(r)$ be the following statement: for every
reachable state $\mathbf M$ consistent with $\mathbf O$ and satisfying
$q(\mathbf M)=r$, the algorithm, when called at $\mathbf M$, returns a subfamily
whose uncovered weight is at most
$(1+\eps)\OPT_{\mathrm{miss}}(I)$.  We prove $P(r)$ by induction on $r$.

Recall that $\delta=\eps/(k(1+\eps))$.  We use the following two claims in
both the base case and the inductive step.

\emph{Claim 1.} At a feasible state $\mathbf M$, the algorithm chooses one set
$A_i\in\mathcal P_i(\mathbf M)$ for each $i\in[k]$.  Before making any
recursive calls, it takes $\mathcal A_{\mathbf M}=\{A_i:i\in[k]\}$ as its
current candidate.  This candidate satisfies
\[
  \miss_{\mathcal S,w}(\mathcal A_{\mathbf M})
  \le w(W(\mathbf M)).
\]

Indeed, $F_i(\mathbf M)$ is the intersection of all sets in
$\mathcal P_i(\mathbf M)$, so $F_i(\mathbf M)\subseteq A_i$ for every $i$.
Thus, $\mathcal A_{\mathbf M}$ covers $F(\mathbf M)$, and only elements in
$W(\mathbf M)=U\setminus F(\mathbf M)$ can remain uncovered.  The algorithm
returns either this candidate or a subfamily with smaller uncovered weight.

\emph{Claim 2.} Let $\mathbf M$ be a reachable state consistent with
$\mathbf O$.  If
$w(W(\mathbf M))>(1+\eps)\OPT_{\mathrm{miss}}(I)$, then some $i\in[k]$ satisfies
$w(O_i\cap W(\mathbf M))>\delta w(W(\mathbf M))$.

To prove the claim, consistency gives
$F_i(\mathbf M)\subseteq O_i$ for every $i$.  Therefore, every element missed
by $\mathbf O$ lies in $W(\mathbf M)$.  These elements have total weight
$\OPT_{\mathrm{miss}}(I)$, so
\[
  w\left(W(\mathbf M)\cap\bigcup_{i=1}^kO_i\right)
  =w(W(\mathbf M))-\OPT_{\mathrm{miss}}(I)
  >\frac{\eps}{1+\eps}w(W(\mathbf M)).
\]
Since the left-hand side is at most
$\sum_{i=1}^k w(O_i\cap W(\mathbf M))$, some $i\in[k]$ satisfies
\[
  w(O_i\cap W(\mathbf M))
  >\frac{\eps}{k(1+\eps)}w(W(\mathbf M))
  =\delta w(W(\mathbf M)),
\]
as required.

For the base case, consider a reachable state $\mathbf M$ consistent with
$\mathbf O$ and satisfying $q(\mathbf M)=0$.  If
$w(W(\mathbf M))\le(1+\eps)\OPT_{\mathrm{miss}}(I)$, Claim~1 proves the
required bound.  Otherwise,
Claim~2 gives an index $i\in[k]$ satisfying the displayed inequality.  Since
$\sum_{h=1}^k|M_h|=kd$ and $|M_h|\le d$ for every $h\in[k]$, we have
$|M_h|=d$ for every $h$.  The inequality above implies that
$O_i\cap W(\mathbf M)\ne\emptyset$ because all weights are positive.  Choose
$u$ from this set and write
$M_i=\{u_1,\ldots,u_d\}$ in insertion order.  By
\cref{lem:semiladder-skeleton-depth}, there are pairwise distinct witness sets
$R_1,\ldots,R_d$.  Since
$u\in W(\mathbf M)\subseteq U\setminus F_i(\mathbf M)$, there is a set
$R_{d+1}\in\mathcal P_i(\mathbf M)$ that misses $u$.  Every set in
$\mathcal P_i(\mathbf M)$ contains $M_i$, so $u_j\in R_{d+1}$ for every
$j\in[d]$.  Moreover, $u\notin M_i$ because
$M_i\subseteq F_i(\mathbf M)$, and, for every $j\in[d]$, we have
$R_{d+1}\ne R_j$ because $u_j\in R_{d+1}\setminus R_j$.  Therefore,
$R_1,\ldots,R_{d+1}$ and
$u_1,\ldots,u_d,u$ form a semi-ladder of order $d+1$, a contradiction.
Thus, $w(W(\mathbf M))>(1+\eps)\OPT_{\mathrm{miss}}(I)$ is impossible, which
proves $P(0)$.

For the inductive step, let $r\in\{1,\ldots,kd\}$ and assume $P(r-1)$.  Let
$\mathbf M$ be a reachable state consistent with $\mathbf O$ and satisfying
$q(\mathbf M)=r$.  If
$w(W(\mathbf M))\le(1+\eps)\OPT_{\mathrm{miss}}(I)$, Claim~1 proves the
required bound.  Otherwise, Claim~2 gives an index $i\in[k]$ such
that $w(O_i\cap W(\mathbf M))>\delta w(W(\mathbf M))$.  Moreover,
$q(\mathbf M)>0$ implies $\sum_{h=1}^k|M_h|<kd$, and
$w(W(\mathbf M))>(1+\eps)\OPT_{\mathrm{miss}}(I)$ implies
$W(\mathbf M)\ne\emptyset$.  The algorithm therefore computes a weighted
$\delta$-net $Z$.  Because
$O_i\in\mathcal P_i(\mathbf M)$, the set $O_i\cap W(\mathbf M)$ belongs to the
restricted family and has weight greater than
$\delta w(W(\mathbf M))$.  Hence $Z$ contains an element
$u\in O_i\cap W(\mathbf M)$.

Let $\mathbf M'$ be obtained by replacing $M_i$ with $M_i\cup\{u\}$.  The set
$O_i$ is eligible at coordinate $i$ and contains $u$, so the algorithm explores
this recursive call.  Moreover, $M_i\subseteq O_i$ and $u\in O_i$, so
$\mathbf M'$ is consistent with $\mathbf O$.
Since $u\in W(\mathbf M)$ and $M_i\subseteq F_i(\mathbf M)$, we have
$u\notin M_i$, and therefore $q(\mathbf M')=q(\mathbf M)-1$.  By the induction
hypothesis, the call at $\mathbf M'$ returns a subfamily of uncovered weight at
most $(1+\eps)\OPT_{\mathrm{miss}}(I)$.  The call at $\mathbf M$ returns a
subfamily of no larger value.  This proves $P(r)$.

The initial state is consistent with $\mathbf O$, so applying the induction
claim to this state proves the lemma.
\end{proof}

\begin{theorem}
\label{thm:semiladder-uncoverage}
For every $d\ge0$, weighted \MinUnCoverage{} on instances whose incidence
graph has semi-ladder index at most $d$ admits a deterministic EPAS
parameterized by $k+d$.  For $0<\eps\le1$, its running time is
\[
  \left(\frac{dk}{\eps}\right)^{
    O\left(
      \frac{d^2k}{\eps}\log\frac{2k}{\eps}
    \right)
  }N^{O(1)},
\]
for $d\ge1$, where $N$ includes the incidence representation and the binary
encoding of the rational weights.  The case $d=0$ is solvable in $O(N)$ time.
\end{theorem}

\begin{proof}
It is enough to consider $0<\eps\le1$.  For $\eps>1$, running the algorithm
with accuracy parameter $1$ gives approximation factor $2\le1+\eps$.

If $k=0$ or $\mathcal S=\emptyset$, return the empty subfamily.  Hence assume
$k\ge1$ and $\mathcal S\ne\emptyset$.  If $d=0$, every set contains every
ground element: a set and an element not contained in it would form a
semi-ladder of order one.  Thus, an arbitrary set is optimal and can be
returned in $O(N)$ time.  We may therefore assume that $d\ge1$.  Correctness
follows from
\cref{lem:semiladder-uncoverage-approximation}.

Recall that $\delta=\eps/(k(1+\eps))$.  By
\cref{lem:semiladder-net}, the weighted $\delta$-net computed at any call has
size
\[
  O\left(\frac{d+1}{\delta}\log\frac1\delta\right).
\]
Since the algorithm considers at most $k$ coordinates for each element of
this weighted $\delta$-net, every call has
\[
  O\left(\frac{k(d+1)}{\delta}\log\frac1\delta\right)
\]
children.

By \cref{lem:semiladder-skeleton-depth}, every skeleton has size at most $d$.
Before making a recursive call, the algorithm adds one new element to exactly
one skeleton.  Hence every
root-to-leaf path has length at most $kd$.  The total number of calls is
therefore
\[
  \left(\frac{k(d+1)}{\delta}\log\frac1\delta\right)^{O(kd)}.
\]

We next bound the work in one call.  The restricted set system is obtained by
restricting eligible input sets to $W(\mathbf M)$, so its encoding size is at
most $N$.  By scanning the incidence lists for each coordinate and each
skeleton element, the algorithm can construct the eligible families, the
forced and unresolved elements, the restricted set system, and the current candidate
using $O(kdN)$ incidence operations.  Computing and comparing candidate
weights takes polynomial time in $N$, since each value is a sum of input
rational weights.  By \cref{lem:semiladder-net}, using the weighted
$\delta$-net construction of Kundu et al.~\cite{kundu2026fixedbudget}, this weighted
$\delta$-net can be computed in
\[
  \left(\frac{d+1}{\delta}\right)^{
    O\left(\frac{(d+1)^2}{\delta}\log\frac1\delta\right)}N^{O(1)}
\]
time.  Thus, the total running time is
\[
  \left(\frac{k(d+1)}{\delta}\log\frac1\delta\right)^{O(kd)}
  \left(\frac{d+1}{\delta}\right)^{
    O\left(\frac{(d+1)^2}{\delta}\log\frac1\delta\right)}N^{O(1)}.
\]

The first factor bounds the number of recursive calls.  Since
$1/\delta\ge2k$, we have
$k\le1/\delta\le(d+1)/\delta$.  Also,
$\log(1/\delta)\le1/\delta\le(d+1)/\delta$.  Thus, all three factors in its
base are at most $(d+1)/\delta$, and
$(k(d+1)/\delta)\log(1/\delta)\le((d+1)/\delta)^3$.  It follows that the
first factor is at most $((d+1)/\delta)^{O(kd)}$.  Moreover,
\[
  \frac{(d+1)^2}{\delta}\log\frac1\delta
  \ge 2(\log 2)kd,
\]
where we use $1/\delta\ge2k$, $\log(1/\delta)\ge\log2$, and
$(d+1)^2\ge d$.  Therefore, the weighted $\delta$-net construction time
absorbs the first factor, and the total running time is
\[
  \left(\frac{d+1}{\delta}\right)^{
    O\left(\frac{(d+1)^2}{\delta}\log\frac1\delta\right)}N^{O(1)}.
\]

Finally, $d+1\le2d$ and $1/\delta\le2k/\eps$.  Therefore,
$(d+1)/\delta\le4dk/\eps$ and
$\log(1/\delta)\le\log(2k/\eps)$.  Substituting these bounds gives
\[
  \left(\frac{dk}{\eps}\right)^{
    O\left(\frac{d^2k}{\eps}\log\frac{2k}{\eps}\right)}N^{O(1)},
\]
which is fixed-parameter tractable in $k+d+1/\eps$.
\end{proof}

By \cref{obs:kdd-semiladder}, every $K_{b,b}$-free incidence graph has
semi-ladder index less than $2b$.  Thus, \cref{thm:semiladder-uncoverage}
also applies to weighted $K_{b,b}$-free instances.  Bounded semi-ladder index
is more general because it permits arbitrarily large bicliques.

%% file: sections/cardinality_augmentation.tex
\section[Approximating the Cardinality Bound (k)]{Approximating the Cardinality Bound ($k$)}
\label{sec:cardinality-augmentation}

This section turns from approximating the uncovered or unsatisfied weight to
relaxing the cardinality bound.  Suppose that a solution using at most $k$
selectable objects leaves weight at most $t$ uncovered or unsatisfied.  Can an
FPT algorithm achieve the same bound using at most $g(k,d)$ objects, for some
computable function $g$?  Because the bound $t$ is fixed, this question is
equivalent in the complementary maximization formulation.

For \MinUnCoverage{}, this is the weighted \textsc{Partial Set Cover} problem;
the relevant positive results and limitations are summarized in
\cref{tab:cardinality-bound}.  For \MinUnsatSAT{} and \MinUnsatCSP{}, we prove
that, unless $\FPT=\Wone$, no FPT algorithm can always attain the same bound
$t$ after replacing $k$ by $g(k,d)$, for any computable function $g$, under
any of the structural restrictions considered here.

\begin{observation}[Exact-cardinality gap]
\label{obs:exact-weight-gap}
Let $I$ be an underlying instance, let $k$ be a cardinality bound, and let
$t$ be a prescribed upper bound on its uncovered or unsatisfied weight.
Suppose that every solution
attaining this bound selects exactly $k$ sets, or, in the SAT/CSP setting,
sets exactly $k$ variables to true.  Then
\[
\OPT(I,k)>t
\quad\Longrightarrow\quad
\OPT(I,q)>t \qquad\text{for every }q\geq 0.
\]
Consequently, increasing the cardinality bound from $k$ to any value
$q$, in particular, to any computable function $g(k,d)$, cannot produce a
solution attaining the prescribed bound.
\end{observation}

\begin{proof}
Suppose instead that $\OPT(I,q)\leq t$ for some $q\geq 0$.  Then there is a
solution whose uncovered or unsatisfied weight is at most $t$.  By assumption,
this solution uses exactly $k$ sets or true variables and is therefore feasible
under the cardinality bound $k$.  Hence $\OPT(I,k)\leq t$, contradicting
$\OPT(I,k)>t$.
\end{proof}

An ordinary bounded-cardinality hardness reduction would not suffice, since a
no-instance under the cardinality bound $k$ might still admit an assignment of larger Hamming
weight whose unsatisfied weight is at most $t$.  Our reductions rule out this
possibility by ensuring that every assignment attaining the prescribed bound
has Hamming weight exactly $k$.  By \cref{obs:exact-weight-gap}, no assignment
of any Hamming weight attains the bound in a no-instance, so replacing $k$ by
any computable function $g(k,d)$ cannot help.

The targets in the constructions below may be positive.  These results concern
preserving a prescribed target and are separate from multiplicative-gap
hardness for approximating a positive optimum value.

\subsection{Hardness under Bounded Arity for \MinUnsatSAT{}}\label{subsec:cardinality-arity}

The EPAS in \cref{sec:bounded-arity} approximates the total weight of
unsatisfied constraints while respecting the cardinality bound.  We
now show that the reverse tradeoff, preserving a bound on the unsatisfied
weight while relaxing the cardinality bound, is impossible even for clauses
of arity three.  The reduction ensures that every assignment attaining the
prescribed bound has Hamming weight exactly $k$.  Consequently, by
\cref{obs:exact-weight-gap}, a no-instance admits no assignment of any Hamming
weight attaining this bound.  We reduce from \textsc{Multicolored Clique} to
\MinUnsatSAT{} as follows.

\paragraph*{Construction.}
Let $(G,V_1,\ldots,V_r)$ be an instance of
\textsc{Multicolored Clique}, where
$V(G)=V_1\uplus\cdots\uplus V_r$ is the partition into color classes, and let
$n=|V(G)|$. Introduce a Boolean variable $x_v$ for every $v\in V(G)$. Call a pair of
distinct vertices $\{u,v\}$ \emph{forbidden} if either $u$ and $v$ belong
to the same color class, or they belong to different color classes and
$uv\notin E(G)$. Thus, a set containing no forbidden pair contains at most
one vertex from each color class and induces a clique.

\begin{itemize}
\item For every vertex $v\in V(G)$, add the unit clause $(x_v)$.

\item Set $W=n+1$. For every forbidden pair $\{u,v\}$ and every
$j\in[W]$, introduce a fresh variable $z_{uv,j}$ and add the gadget clauses
$(\neg x_u\vee\neg x_v\vee z_{uv,j})$ and $(\neg z_{uv,j})$.
\item  Set $k=r$ and $t=n-r$.
\end{itemize}

Let $\Phi_G$ denote the resulting formula.  Every clause has arity at most
three.
\begin{claim}
\label{clm:arity-equivalence}
 $G$ contains a multicolored clique of size $r$ if and only if
$\Phi_G$ has an assignment leaving at most $t$ clauses unsatisfied.
\end{claim}

\begin{proof}
($\Rightarrow$) Suppose first that $G$ contains a multicolored clique
$S=\{v_1,\ldots,v_r\}$, where $v_i\in V_i$. Set $x_v=1$ exactly for
$v\in S$ and set every fresh variable $z_{uv,j}$ to zero. Since $S$
contains no forbidden pair, every clause involving a fresh variable is
satisfied. Among the unit clauses $(x_v)$, exactly those corresponding to
vertices outside $S$ are unsatisfied. Hence the assignment leaves exactly
$n-r=t$ clauses unsatisfied and has Hamming weight exactly $r=k$.
\smallskip

($\Leftarrow$) Suppose that $\tau$ is an assignment to $\Phi_G$ leaving at
most $t=n-r$ clauses unsatisfied, and let
$S=\{v\in V(G):\tau(x_v)=1\}$. We first claim that $S$ contains no
forbidden pair. Indeed, suppose that $\{u,v\}\subseteq S$ is forbidden.
Then, for every $j\in[W]$, the two corresponding clauses reduce to
$(z_{uv,j})$ and $(\neg z_{uv,j})$, so at least one of them is unsatisfied.
Thus $\tau$ leaves at least $W=n+1>n-r=t$ clauses unsatisfied, a
contradiction. Hence $S$ contains no forbidden pair. Therefore, $S$ contains at most one
vertex from each color class and induces a clique, so $|S|\le r$. On the
other hand, the unit clauses $(x_v)$ alone contribute exactly $n-|S|$
unsatisfied clauses. Since $\tau$ leaves at most $n-r$ clauses unsatisfied,
we have $n-|S|\le n-r$, and hence $|S|\ge r$. Thus $|S|=r$, so $S$
contains exactly one vertex from each color class and is a multicolored
clique.
\end{proof}
\begin{theorem}
\label{thm:cardinality-arity}
Unless $\FPT=\Wone$, for every computable function
$g\colon\mathbb N^2\to\mathbb N$ satisfying $g(k,d)\ge k$, there is no
$f(k,d)N^{O(1)}$-time algorithm for unweighted \MinUnsatSAT{} with clause
arity at most $d$ that, whenever $\OPT(I,k)\le t$, returns an assignment of
Hamming weight at most $g(k,d)$ and objective value at most $t$.
In particular, no finite-factor approximation of the cardinality bound is
possible in FPT time parameterized by $k$ while preserving the objective
target, even for $d=3$.
\end{theorem}

\begin{proof}
By \cref{clm:arity-equivalence}, $G$ has a multicolored clique if and only if
$\Phi_G$ has an assignment leaving at most $t=n-r$ clauses unsatisfied.  The
forward direction gives such an assignment of Hamming weight exactly $r=k$.
Conversely, the proof of the claim shows that every assignment attaining this
bound sets exactly $r$ original variables to one.  The corresponding $n-r=t$
unsatisfied unit clauses exhaust the bound, so every gadget clause is
satisfied.  In particular, every clause $(\neg z_{uv,j})$ is satisfied, and
hence every fresh variable is zero.  Thus every assignment attaining the
bound has Hamming weight exactly $k$.

The hypotheses of \cref{obs:exact-weight-gap} are therefore satisfied.  On a
yes-instance an assignment of Hamming weight $k$ attains the bound, whereas on
a no-instance no assignment of any Hamming weight does.  An algorithm with the
guarantee in the theorem would consequently decide \textsc{Multicolored
Clique} in FPT time by running it on the constructed instance and checking
whether its output leaves at most $t$ clauses unsatisfied.  Since $k=r$ and
$d=3$, this is a parameterized reduction, completing the proof.
\end{proof}

\subsection{Hardness for Bounded-Variable-Frequency \MinUnsatSAT{}}
\label{subsec:cardinality-frequency}
For bounded variable frequency, we show that approximating the cardinality
bound is again impossible. As discussed after \cref{obs:exact-weight-gap}, we
amplify the cost of leaving a clause unsatisfied so that even an assignment of
unrestricted Hamming weight cannot attain the target in the no-case. For our
purpose, we reduce from an instance of \textsc{Multicolored Clique} to an
equivalent instance of \MinUnsatSAT{} as follows.
\paragraph*{Construction.}
Let $(G,V_1,\ldots,V_r)$ be an instance of \textsc{Multicolored Clique}.
If $E(V_i,V_j)=\emptyset$ for some pair of distinct color classes, then $G$ is
a no-instance; in this case we return the fixed formula
$(z)\wedge(\neg z)$ with cardinality bound and objective target both zero.
We may therefore assume that every pair of color classes has at least one edge.

We first apply the bounded-variable-frequency construction from
\cref{sec:kdd-hardness}.  By
\cref{clm:frequency-bound,clm:frequency-equivalence}, it produces a formula $\Phi_G$
and a cardinality bound $K=r+\binom r2$ such that $G$ contains a
multicolored clique if and only if $\Phi_G$ has a satisfying assignment of
Hamming weight at most $K$. Moreover, every variable of $\Phi_G$ has
frequency at most $d_0=\max\{r,3\}$.

We construct a new formula $\Psi_G$ as follows.

\begin{itemize}
\item Set $W=K+1$. For every clause $C$ of $\Phi_G$ and every $j\in[W]$,
introduce a fresh variable $z_{C,j}$ and add the clauses
$(C\vee z_{C,j})$ and $(\neg z_{C,j})$.

\item For every original variable $x$ of $\Phi_G$, add the unit clause
$(\neg x)$.

\item Set $k=K$ and $t=K$.
\end{itemize}

Note that every original variable has frequency at most $Wd_0+1$, while
every fresh variable has frequency exactly two.

\begin{claim}
\label{clm:frequency-cardinality-equivalence}
$G$ contains a multicolored clique of size $r$ if and only if $\Psi_G$ has
an assignment leaving at most $t$ clauses unsatisfied.
\end{claim}

\begin{proof}
($\Rightarrow$) Suppose first that $G$ contains a multicolored clique. By the
forward direction of \cref{clm:frequency-equivalence}, $\Phi_G$ has a
satisfying assignment of Hamming weight at most $K$.  The formula contains
exactly $K$ selection clauses with pairwise disjoint variable sets, so this
assignment has Hamming weight exactly $K$. Extend it by setting
every fresh variable $z_{C,j}$ to zero. Since every clause $C$ of $\Phi_G$
is satisfied, both $(C\vee z_{C,j})$ and $(\neg z_{C,j})$ are satisfied for
every $C$ and $j$. Among the unit clauses $(\neg x)$, exactly those
corresponding to the $K$ original variables set to one are unsatisfied.
Hence the assignment leaves exactly $K=t$ clauses unsatisfied and has
Hamming weight exactly $K=k$.

($\Leftarrow$) Suppose that $\tau$ is an assignment to $\Psi_G$ leaving at
most $t=K$ clauses unsatisfied. We first claim that the restriction of
$\tau$ to the original variables satisfies every clause of $\Phi_G$.
Indeed, suppose that some clause $C$ of $\Phi_G$ is unsatisfied. Then, for
every $j\in[W]$, the two corresponding clauses reduce to $(z_{C,j})$ and
$(\neg z_{C,j})$, so at least one of them is unsatisfied. Thus $\tau$ leaves
at least $W=K+1>t$ clauses unsatisfied, a contradiction.

Hence the restriction of $\tau$ satisfies $\Phi_G$. Let $s$ be the number
of original variables set to one. The unit clauses $(\neg x)$ alone
contribute exactly $s$ unsatisfied clauses. Since $\tau$ leaves at most
$K$ clauses unsatisfied, we have $s\le K$. Thus $\Phi_G$ has a satisfying
assignment of Hamming weight at most $K$, and hence the reverse direction of
\cref{clm:frequency-equivalence} implies that $G$ contains a
multicolored clique.
\end{proof}

\begin{theorem}
\label{thm:cardinality-frequency}
Unless $\FPT=\Wone$, for every computable function
$g\colon\mathbb N^2\to\mathbb N$ satisfying $g(k,d)\ge k$, there is no
$f(k,d)N^{O(1)}$-time algorithm for unweighted \MinUnsatSAT{} with maximum
variable frequency at most $d$ that, whenever $\OPT(I,k)\le t$, returns an
assignment of Hamming weight at most $g(k,d)$ and objective value at most
$t$. In particular, no finite-factor approximation of the cardinality bound
is possible in FPT time parameterized by $k+d$ while preserving the objective
target.
\end{theorem}

\begin{proof}
By \cref{clm:frequency-cardinality-equivalence}, $G$ has a multicolored clique
if and only if $\Psi_G$ has an assignment leaving at most $t=K$ clauses
unsatisfied.  In the forward direction, the constructed assignment has
Hamming weight exactly $K=k$.

For the reverse direction, recall that $\Phi_G$ contains exactly $K$
selection clauses with pairwise disjoint variable sets.  Every satisfying
assignment of $\Phi_G$ therefore sets at least $K$ original variables to one.
The proof of \cref{clm:frequency-cardinality-equivalence} shows that an
assignment to $\Psi_G$ attaining the bound restricts to a satisfying
assignment of $\Phi_G$ with at most $K$ true variables.  It consequently has
exactly $K$ true original variables.  The corresponding $K=t$ unit clauses
$(\neg x)$ exhaust the bound, so all remaining clauses are satisfied.  In
particular, every clause $(\neg z_{C,j})$ is satisfied, and hence all fresh
variables are zero.  Thus every assignment attaining the bound has Hamming
weight exactly $K=k$.

By \cref{obs:exact-weight-gap}, a no-instance admits no assignment of any
Hamming weight attaining the bound.  Therefore, an algorithm with the
guarantee in the theorem would decide \textsc{Multicolored Clique} in FPT time.
Finally, the maximum variable frequency of $\Psi_G$ satisfies
$d\leq (K+1)d_0+1=O(r^3)$, while $k=K=O(r^2)$, so the output parameter $k+d$
is bounded by a function of $r$.
\end{proof}

The two preceding constructions yield hardness under all the remaining
structural restrictions.

\begin{corollary}
\label{cor:cardinality-other-restrictions}
Unless $\FPT=\Wone$, for every computable function
$g\colon\mathbb N^2\to\mathbb N$ satisfying $g(k,d)\geq k$, unweighted
\MinUnsatSAT{} admits no
$f(k,d)N^{O(1)}$-time algorithm that, whenever $\OPT(I,k)\leq t$, returns an
assignment of Hamming weight at most $g(k,d)$ and unsatisfied weight at most
$t$, under any of the following restrictions:
\begin{enumerate}
    \item bounded clause or constraint arity;
    \item bounded variable frequency;
    \item $K_{d,d}$-free incidence graphs;
    \item bounded semi-ladder index; or
    \item bounded VC dimension.
\end{enumerate}
The same lower bounds hold for unweighted \MinUnsatCSP{}.
The arity case already holds for $d=3$.
For the last three restrictions, the hardness already holds for
$K_{3,3}$-free incidence graphs, semi-ladder index at most four, and
VC dimension at most two, respectively.
\end{corollary}

\begin{proof}
The bounded-variable-frequency case is
\cref{thm:cardinality-frequency}.  All the other claims follow from the
arity-three construction in \cref{thm:cardinality-arity}, which already has
the required exact-cardinality gap.

Its incidence graph is $K_{3,3}$-free.  Indeed, every fresh variable
$z_{uv,j}$ has degree two and hence cannot lie on the variable side of a
$K_{3,3}$.  The remaining variables are the original variables $x_v$, but
every clause contains at most two of them, so three such variables cannot
have a common clause neighbor.

The clause-degree bound also gives the other two restrictions.  A semi-ladder
of order $\ell$ contains a clause adjacent to at least $\ell-1$ variables, so
$\ell\leq4$.  If $\ell$ clauses are shattered by variable neighborhoods, then
each of them has at least $2^{\ell-1}$ distinct variable neighbors.  Since
every clause has degree at most three, $2^{\ell-1}\leq3$ and hence
$\ell\leq2$.

Finally, every CNF formula is a Boolean CSP instance under the input convention
of \cref{sec:preliminaries}, with the cardinality bound, unsatisfied weight,
and incidence graph unchanged.
\end{proof}

Thus the additive-one guarantee for coverage on incidence graphs of bounded
semi-ladder index contrasts sharply with SAT and Boolean CSP, for which no
computable relaxation of the cardinality bound is possible even under bounded
arity or bounded variable frequency.

%% file: sections/conclusion.tex
\section{Conclusion}
\label{sec:conclusion}

We studied the parameterized approximability of the residual-minimization
counterparts of cardinality-constrained coverage and satisfaction problems.
Our results reveal a sharp structural distinction between coverage and general
satisfaction.  Bounded arity yields a deterministic EPAS for weighted
\MinUnsatCSP{}.  Via the monotone-CNF encoding, this also gives a deterministic
EPAS for weighted \MinUnCoverage{} when every ground element has bounded
frequency.  Bounded set size additionally yields an exact FPT algorithm for
weighted \MinUnCoverage{}.  For broader restrictions on the incidence graph,
bounded semi-ladder index supports a deterministic EPAS for weighted
\MinUnCoverage{}, whereas \MinUnsatSAT{} and \MinUnsatCSP{} admit no
finite-factor FPT approximation even on unweighted $K_{3,3}$-free instances
and on instances of semi-ladder index at most four.  Bounded VC dimension
alone does not permit a finite-factor FPT approximation for any of the three
problems.

We also studied relaxing the cardinality bound while maintaining a prescribed
upper bound on the uncovered or unsatisfied weight.  For weighted
\textsc{Partial Set Cover}, bounded semi-ladder index allows an algorithm to
meet the same bound using at most $k+1$ sets whenever it can be met using $k$
sets.  For unweighted \MinUnsatSAT{} and \MinUnsatCSP{}, in contrast, no FPT
algorithm can preserve the bound under any computable relaxation of the
cardinality constraint.  This hardness holds under every structural
restriction considered here, already for arity three, bounded variable
frequency, $K_{3,3}$-free incidence graphs, semi-ladder index at most four,
and VC dimension at most two.  Thus the distinction between coverage and
general satisfaction persists both for multiplicative approximation of the
residual weight and for relaxation of the cardinality bound.

\paragraph*{AI disclosure.}
OpenAI Codex and Anthropic Claude Opus were used to reorganize and copy-edit
the manuscript, formalize the exposition of author-supplied arguments, check
proofs for typographical errors and minor logical inconsistencies, suggest
clearer phrasing, and draft selected proof passages. The authors are
responsible for verifying all mathematical claims, proofs, citations, and
the final text.

%% file: sections/min_uncoverage.tex
\section[An Improved EPAS for Unit-Weight Biclique-Free \MinUnCoverage{}]
{An Improved EPAS for Unit-Weight \\
$K_{d,d}$-Free \MinUnCoverage{}}
\label{sec:min-uncoverage}

The EPAS for weighted \MinUnCoverage{} in
\cref{thm:semiladder-uncoverage} applies to
$K_{d,d}$-free instances because their semi-ladder index is less than $2d$.
For unit weights, the algorithm in this appendix has better parameter
dependence.  Throughout this section, every ground element has weight one.
Fix $d\ge2$.  We describe the algorithm for a current $K_{d,d}$-free instance
$I=(U,\mathcal S,q)$, where $q$ is the number of sets that may still be
selected.  The initial instance has $q=k$.  Write $m=|U|$, and let $N$ denote
the input encoding size.  Define
\[
  \OPT_{\mathrm{miss}}(I)
  =\min_{\substack{\mathcal A\subseteq\mathcal S\\|\mathcal A|\le q}}
    \miss_{\mathcal S}(\mathcal A)
  \quad\text{and}\quad
  \OPT_{\mathrm{cov}}(I)
  =\max_{\substack{\mathcal A\subseteq\mathcal S\\|\mathcal A|\le q}}
    \cov_{\mathcal S}(\mathcal A).
\]
Thus $\OPT_{\mathrm{miss}}(I)+\OPT_{\mathrm{cov}}(I)=m$.

If $q=0$ or $U$ is empty, the algorithm stops.  It also stops and solves the
instance exactly when $m\le4dq$.  In every other call, it computes two types
of solutions.  First, it runs a deterministic
approximation scheme for \textsc{Maximum Coverage}.  This gives the required
$(1+\eps)$-approximation when an optimum solution leaves at least half of $U$
uncovered.

Now suppose that an optimum solution leaves fewer than $m/2$ elements
uncovered.  It then covers more than $m/2$ elements with at most $q$ sets, so
one of its selected sets contains at least $m/(2q)$ elements.  The algorithm
tries every set $S$ of at least this size.  In the branch for $S$, it selects
$S$, removes the elements covered by $S$, and solves the remaining instance
recursively with budget $q-1$.  It then lifts the recursive solution back to
sets of the current instance.  It returns the solution that leaves the fewest
elements uncovered.

We first bound the number of large sets that the algorithm may have to try.

\begin{lemma}[Number of large sets]
\label{lem:kdd-high-degree}
Let $(U,\mathcal S)$ be a $K_{d,d}$-free set system, where $|U|=m$, and let
$t\ge2d$ be an integer.  Then
\[
  \bigl|\{S\in\mathcal S:|S|\ge t\}\bigr|
  \le (d-1)\left(\frac{2m}{t}\right)^d.
\]
\end{lemma}

\begin{proof}
If $t>m$, there are no such sets.  Assume that $t\le m$, and let
$H_t=\{S\in\mathcal S:|S|\ge t\}$.  Count pairs $(S,D)$ such that
$S\in H_t$ and $D$ is a $d$-element subset of $S$.  Every set in $H_t$
contributes at least $\binom{t}{d}$ pairs.  Each $d$-element subset of $U$ is
contained in at most $d-1$ sets, because $d$ sets containing the same $d$
elements would form a $K_{d,d}$.  Therefore,
\[
  |H_t|\binom{t}{d}\le (d-1)\binom{m}{d}.
\]
Because $t\ge2d$, we have $t-d+1\ge t/2$.  Moreover, for every
$i\in\{0,\ldots,d-1\}$, we have $m-i\le m$ and $t-i\ge t-d+1$.  Hence
\[
  \frac{\binom{m}{d}}{\binom{t}{d}}
  =\prod_{i=0}^{d-1}\frac{m-i}{t-i}
  \le\left(\frac{m}{t-d+1}\right)^d
  \le\left(\frac{2m}{t}\right)^d.
\]
Combining this inequality with the preceding counting bound proves the lemma.
\end{proof}

The second tool solves the instance exactly when the ground set is small.

\begin{lemma}[Exact solution for a small ground set]
\label{lem:small-family}
Given an instance $I=(U,\mathcal S,q)$ with $m$ ground elements, an optimum
solution can be found in $2^{O(m)}N^{O(1)}$ time.
\end{lemma}

\begin{proof}
For each $t\in[m]$, the deterministic version of Bl\"aser's color-coding
algorithm finds a minimum-cardinality family covering at least $t$ elements in
$2^{O(t)}N^{O(1)}$ time~\cite{DBLP:journals/ipl/Blaser03,alon1995colorcoding}.
Run it for every $t$ and return the solution for the largest $t$ achievable
with at most $q$ sets, or the empty family if no positive $t$ is achievable.
Since $t\le m$, all $m$ calls together take $2^{O(m)}N^{O(1)}$ time.
\end{proof}

The third tool is the deterministic EPAS for biclique-free \textsc{Maximum
Coverage}.  Jain et al.\ established an EPAS directly for \textsc{Maximum
Coverage}~\cite{jain2023parameterized}.  The improved running time used below
follows by applying Manurangsi's later deterministic scheme for biclique-free
\textsc{Max $k$-Weight SAT} to the monotone-CNF
encoding~\cite{manurangsi2025improved}.

\begin{proposition}
\label{prop:kdd-max-coverage-epas}
Let $d\ge2$ and $0<\eps\le1$.  Given an unweighted $K_{d,d}$-free instance
$I=(U,\mathcal S,q)$, a deterministic algorithm returns a subfamily
$\mathcal A_{\mathrm{cov}}\subseteq\mathcal S$ with $|\mathcal A_{\mathrm{cov}}|\le q$
such that
\[
  \cov_{\mathcal S}(\mathcal A_{\mathrm{cov}})
  \ge (1-\eps)\OPT_{\mathrm{cov}}(I).
\]
Its running time is $(dq/\eps)^{O(dq)}N^{O(1)}$.
\end{proposition}

We now combine
\cref{lem:kdd-high-degree,lem:small-family,prop:kdd-max-coverage-epas}
in the recursive approximation algorithm.

\paragraph*{Residual instances.}
Consider a current instance $I=(U,\mathcal S,q)$ with $q\ge1$.  Define the
branching family
$H(I)=\{S\in\mathcal S:|S|\ge m/(2q)\}$.  We next define the instance used
after the algorithm selects a set $S\in H(I)$.

Put $U_S:=U\setminus S$ and
\[
  \mathcal S_S:=\{T\setminus S:T\in\mathcal S\setminus\{S\}\}.
\]
If several sets of the current instance give the same member $R$ of
$\mathcal S_S$, retain only one copy and fix a representative
$\rho_S(R)\in\mathcal S\setminus\{S\}$ satisfying
$\rho_S(R)\setminus S=R$.  For
$\mathcal B\subseteq\mathcal S_S$, write
$\rho_S(\mathcal B):=\{\rho_S(R):R\in\mathcal B\}$.  Distinct residual sets
have distinct representatives, so
$|\rho_S(\mathcal B)|=|\mathcal B|$.  Define
\[
  I\mathbin{\downarrow}S:=(U_S,\mathcal S_S,q-1).
\]
Its incidence graph is isomorphic to a subgraph of the incidence graph of $I$:
delete the vertex $S$, the element vertices in $S$, and any set
vertices not chosen as representatives.  Hence the residual instance remains
$K_{d,d}$-free.
We write $\miss_{I\mathbin{\downarrow}S}(\mathcal B)$ for the number of
elements in the new ground set that $\mathcal B$ leaves uncovered.  For every
$\mathcal B\subseteq\mathcal S_S$,
\begin{equation}
\label{eq:coverage-residual}
  \miss_{\mathcal S}(\rho_S(\mathcal B)\cup\{S\})
  =\miss_{I\mathbin{\downarrow}S}(\mathcal B).
\end{equation}

\paragraph*{The recursive algorithm.}
The full procedure is given in \cref{alg:min-uncoverage}.  If $q=0$ or $U$ is
empty, it returns the empty family.  If $m\le4dq$, it uses the exact algorithm from
\cref{lem:small-family}.  Otherwise, $m/(2q)>2d$, so
\cref{lem:kdd-high-degree} bounds the size of $H(I)$.  The algorithm computes
$\mathcal A_{\mathrm{cov}}$ using \cref{prop:kdd-max-coverage-epas}, recursively
tries every set in $H(I)$, and returns the solution that leaves the fewest
elements uncovered.

\begin{algorithm}[H]
\caption{$\textsc{ApxMinUnCov}(I,d,\eps)$}
\label{alg:min-uncoverage}
\begin{algorithmic}[1]
\Input An unweighted $K_{d,d}$-free instance $I=(U,\mathcal S,q)$ and
$\eps\in(0,1]$.
\Output A subfamily $\mathcal A\subseteq\mathcal S$ with
$|\mathcal A|\le q$.
\If{$q=0$ or $U=\emptyset$}
  \State \Return $\emptyset$
\EndIf
\If{$|U|\le 4dq$}
  \State \Return an optimum solution for budget $q$ computed using
    \cref{lem:small-family}
\EndIf
\State Compute $\mathcal A_{\mathrm{cov}}$ for budget $q$ using
\cref{prop:kdd-max-coverage-epas}
\State $H\gets\{S\in\mathcal S:|S|\ge |U|/(2q)\}$
\State $\mathcal A_{\mathrm{rec}}\gets\emptyset$
\ForAll{$S\in H$}
  \State $\mathcal B_S\gets
    \textsc{ApxMinUnCov}(I\mathbin{\downarrow}S,d,\eps)$
  \State $\mathcal A_S\gets\{S\}\cup\rho_S(\mathcal B_S)$
  \If{$\miss_{\mathcal S}(\mathcal A_S)
      <\miss_{\mathcal S}(\mathcal A_{\mathrm{rec}})$}
    \State $\mathcal A_{\mathrm{rec}}\gets\mathcal A_S$
  \EndIf
\EndFor
\State \Return whichever of $\mathcal A_{\mathrm{cov}}$ and
$\mathcal A_{\mathrm{rec}}$ leaves fewer elements uncovered
\end{algorithmic}
\end{algorithm}

We first prove the approximation guarantee.

\begin{lemma}[Approximation guarantee]
\label{lem:coverage-approximation}
Let $d\ge2$, let $0<\eps\le1$, and let $I=(U,\mathcal S,q)$ be an unweighted
$K_{d,d}$-free instance.  Then \cref{alg:min-uncoverage} returns at most $q$
sets that leave at most $(1+\eps)\OPT_{\mathrm{miss}}(I)$ ground elements
uncovered.
\end{lemma}

\begin{proof}
Fix $d\ge2$ and $0<\eps\le1$.  For $q\ge0$, let $P(q)$ be the following
statement: for every unweighted $K_{d,d}$-free instance
$I=(U,\mathcal S,q)$, the call $\textsc{ApxMinUnCov}(I,d,\eps)$ returns at most
$q$ sets and leaves at most $(1+\eps)\OPT_{\mathrm{miss}}(I)$ elements
uncovered.  We prove $P(q)$ by induction on $q$.

For the base case, let $q=0$.  The algorithm returns the empty family, which
is the only feasible solution.  Thus it is optimal, and $P(0)$ holds.

For the inductive step, let $q\ge1$ and assume $P(q-1)$.  Fix an unweighted
$K_{d,d}$-free instance $I=(U,\mathcal S,q)$, and write $m=|U|$.  If $U$ is
empty or $m\le4dq$, the algorithm returns an optimum solution using at most
$q$ sets.  Hence assume that $m>4dq$.

Every candidate considered by the algorithm uses at most $q$ sets.  By
\cref{prop:kdd-max-coverage-epas}, the candidate
$\mathcal A_{\mathrm{cov}}$ uses at most $q$ sets.
For each $S\in H(I)$, the residual instance $I\mathbin{\downarrow}S$ remains
$K_{d,d}$-free.  By the induction hypothesis, the recursive call returns at
most $q-1$ residual sets.  Since $\rho_S$ preserves cardinality,
$\mathcal A_S=\{S\}\cup\rho_S(\mathcal B_S)$ uses at most $q$ sets.
Therefore, the family returned by the algorithm is feasible.

We now prove its approximation guarantee.  Suppose first that
$\OPT_{\mathrm{miss}}(I)\ge m/2$.  Then
$\OPT_{\mathrm{cov}}(I)=m-\OPT_{\mathrm{miss}}(I)$ is at most
$\OPT_{\mathrm{miss}}(I)$.  The guarantee in
\cref{prop:kdd-max-coverage-epas} gives
\begin{align*}
  \miss_{\mathcal S}(\mathcal A_{\mathrm{cov}})
  &=m-\cov_{\mathcal S}(\mathcal A_{\mathrm{cov}})\\
  &\le m-(1-\eps)\OPT_{\mathrm{cov}}(I)\\
  &=\OPT_{\mathrm{miss}}(I)+\eps\OPT_{\mathrm{cov}}(I)\\
  &\le(1+\eps)\OPT_{\mathrm{miss}}(I).
\end{align*}
Thus $\mathcal A_{\mathrm{cov}}$, and therefore the family returned by the
algorithm, has the required value.

Now suppose that $\OPT_{\mathrm{miss}}(I)<m/2$, and let $\mathcal A^*$ be
an optimum solution.  It covers more than $m/2$ elements using at most $q$
sets.  Therefore, some $S^*\in\mathcal A^*$ contains at least $m/(2q)$
elements.  Hence $S^*\in H(I)$, and the algorithm tries the recursive choice
that selects $S^*$.

Define
\[
  \mathcal B^*:=
  \{T\setminus S^*:T\in\mathcal A^*\setminus\{S^*\}\}
  \subseteq\mathcal S_{S^*}.
\]
Merging duplicate residual sets can only decrease cardinality, so
$|\mathcal B^*|\le q-1$.  Moreover, $\mathcal B^*$ leaves exactly the elements
left uncovered by $\mathcal A^*$, and hence
$\OPT_{\mathrm{miss}}(I\mathbin{\downarrow}S^*)\le
\OPT_{\mathrm{miss}}(I)$.  Conversely, every solution $\mathcal B$ for the
residual instance lifts to
$\rho_{S^*}(\mathcal B)\cup\{S^*\}$, which has the same number of uncovered
elements by \cref{eq:coverage-residual}.  Therefore,
$\OPT_{\mathrm{miss}}(I\mathbin{\downarrow}S^*)
=\OPT_{\mathrm{miss}}(I)$.

The residual instance is $K_{d,d}$-free, so the induction hypothesis applies.
It returns a solution $\mathcal B_{S^*}$ for the residual instance that leaves
at most $(1+\eps)\OPT_{\mathrm{miss}}(I)$ elements uncovered.  The lifted
solution $\rho_{S^*}(\mathcal B_{S^*})\cup\{S^*\}$ preserves that number by
\cref{eq:coverage-residual}.  The algorithm returns a solution with no larger
value.  This proves $P(q)$ and completes the induction.
\end{proof}

We can now prove the main result.

\begin{theorem}
\label{thm:kdd-coverage}
For every $d\ge2$, unweighted $K_{d,d}$-free \MinUnCoverage{} admits an EPAS
parameterized by $k+d$.  The scheme is deterministic and runs in
$(dk/\eps)^{O(dk)}N^{O(1)}$ time for $0<\eps\le1$, where $N$ is the input
encoding size.
\end{theorem}

\begin{proof}
It is enough to consider $0<\eps\le1$.  If $\eps>1$, we run the algorithm with
accuracy parameter $1$; the resulting approximation factor of $2$ is at most
$1+\eps$.  Run
$\textsc{ApxMinUnCov}(I,d,\eps)$ on the input instance
$I=(U,\mathcal S,k)$.  The approximation guarantee follows from
\cref{lem:coverage-approximation}.

If $k=0$, the algorithm stops immediately.  Assume therefore that $k\ge1$.
We now bound the number of recursive calls.  Consider a call on a current
instance $J=(U',\mathcal S',q)$ that does not stop, and let $m=|U'|$.  Since
$m>4dq$, the integer $t=\lceil m/(2q)\rceil$ is at least $2d$.  The family
$H(J)$ consists
exactly of the sets of size at least $t$.  By \cref{lem:kdd-high-degree},
\[
  |H(J)|
  \le(d-1)\left(\frac{2m}{t}\right)^d
  \le(d-1)(4q)^d
  \le(d-1)(4k)^d.
\]
Each recursive call reduces the remaining budget by one, so there are at most
$k$ recursive levels.  Let $B=(d-1)(4k)^d$.  The total number of calls is at
most $1+B+\cdots+B^k\le(1+B)^k$.

Every recursive instance, including its representative map, has encoding size
$O(N)$.  By \cref{prop:kdd-max-coverage-epas}, the \textsc{Maximum Coverage}
computation at each nonterminal call therefore takes at most
$(dk/\eps)^{O(dk)}N^{O(1)}$ time.  When the algorithm stops because the ground
set is small, it has at most $4dq\le4dk$ elements.  By
\cref{lem:small-family}, solving such an instance takes
$2^{O(dk)}N^{O(1)}$ time.  Constructing branching families and residual
instances, and comparing the solutions, takes polynomial time.  Therefore,
the total running time is at most
\[
  \bigl(1+(d-1)(4k)^d\bigr)^k
  \cdot\bigl((dk/\eps)^{O(dk)}+2^{O(dk)}\bigr)
  \cdot N^{O(1)}
  =(dk/\eps)^{O(dk)}N^{O(1)},
\]
which is fixed-parameter tractable in $k+d+1/\eps$.
\end{proof}

For comparison, the unit-weight algorithm of \cref{thm:kdd-coverage} runs in
$(dk/\eps)^{O(dk)}N^{O(1)}$ time.  By \cref{obs:kdd-semiladder}, every
$K_{d,d}$-free incidence graph has semi-ladder index less than $2d$.  Hence
\cref{thm:semiladder-uncoverage} gives, for weighted $K_{d,d}$-free instances,
running time
\[
  \left(\frac{dk}{\eps}\right)^{
    O\left(
      \frac{d^2k}{\eps}\log\frac{2k}{\eps}
    \right)
  }N^{O(1)}.
\]
Thus the unit-weight algorithm has better parameter dependence, whereas the
weighted algorithm accommodates positive rational weights
and applies more generally to incidence graphs of bounded semi-ladder index.

%% file: sections/bounded_set_size.tex
\section{An Exact Algorithm for Weighted \MinUnCoverage{} with Bounded Set Size}
\label{sec:bounded-set-size}

For unweighted \textsc{Maximum Coverage}, fixed-parameter tractability under the
combined parameter consisting of the number of selected sets and the maximum
set size follows from Bl\"aser's randomized color-coding algorithm for
\textsc{Partial Set Cover}: a family of at most $k$ sets of size at most $d$
covers at most $kd$ elements~\cite{DBLP:journals/ipl/Blaser03}.  Bonnet et al.\
gave an alternative deterministic FPT algorithm based on
greediness-for-parameterization~\cite{bonnet2016maxksetcover}.  We adapt
Bl\"aser's color-coding framework to positive rational element weights and
make the adaptation deterministic.  The proof appears here because it closely
follows that framework; the substantive modifications are the treatment of
weights and the standard derandomization.  This monotone setting is essential:
cardinality-constrained
\textsc{Red--Blue Set Cover} is $\Wone$-hard parameterized by the number of
selected sets plus the red budget, even when every set has size
three~\cite{ashok2017redblue}.

\begin{proposition}
\label{prop:bounded-set-size-coverage}
Weighted \MinUnCoverage{} on instances in which every set has size at most
$d$ can be solved exactly by a deterministic algorithm in
$2^{O(kd)}N^{O(1)}$ time.
\end{proposition}

\begin{proof}
Let $I=(U,\mathcal S,w,k)$ be such an instance.  Since
\[
  \miss_{\mathcal S,w}(\mathcal A)
  =w(U)-w\!\left(\bigcup_{S\in\mathcal A}S\right),
\]
it suffices to maximize the covered weight.  Denote this maximum, over
subfamilies of size at most $k$, by $\OPT_{\mathrm{cov}}(I)$.  The cases
$k=0$ and $d=0$ are immediate, so assume $k,d\ge1$ and put $p:=kd$.

\paragraph*{An equivalent disjoint-subset formulation.}
For every $S\in\mathcal S$ and every nonempty $T\subseteq S$, form the ordered
pair $(S,T)$.  Its first component records the set that will be selected, and
its second component records the elements whose weight is counted for that
pair.  Let
\[
  \mathcal P:=\{(S,T):S\in\mathcal S,\ \emptyset\ne T\subseteq S\},
\]
and give $(S,T)$ value $w(T)$.  Since $|S|\le d$,
\[
  |\mathcal P|\le(2^d-1)|\mathcal S|.
\]
Call a collection of at most $k$ pairs from $\mathcal P$ \emph{feasible} if
their second components are pairwise disjoint, and let
$\OPT_{\mathrm{pack}}(I)$ be the maximum total value of a feasible collection.

We claim that
\[
  \OPT_{\mathrm{cov}}(I)=\OPT_{\mathrm{pack}}(I).
\]
Indeed, order any coverage solution $S_1,\ldots,S_q$, where $q\le k$, and set
\[
  T_i:=S_i\setminus\bigcup_{j<i}S_j.
\]
After discarding the empty $T_i$, the resulting pairs $(S_i,T_i)$ are feasible
and have total value exactly $w(\bigcup_{i=1}^qS_i)$.  This proves
$\OPT_{\mathrm{cov}}(I)\le\OPT_{\mathrm{pack}}(I)$.  Conversely, from any
feasible pairs $(S_1,T_1),\ldots,(S_q,T_q)$, select the distinct sets among
$S_1,\ldots,S_q$.  There are at most $q\le k$ of them, and they cover
$\bigcup_iT_i$.  Positivity of the weights and pairwise disjointness give
\[
  w\!\left(\bigcup_{i=1}^qS_i\right)
  \ge w\!\left(\bigcup_{i=1}^qT_i\right)
  =\sum_{i=1}^q w(T_i),
\]
which proves the reverse inequality.  This conversion also recovers a
coverage solution from feasible pairs in polynomial time.

\paragraph*{Color coding and the dynamic program.}
A coloring is a function $h\colon U\to[p]$; we say that $h$ is \emph{good}
for $X\subseteq U$ if it assigns distinct colors to distinct elements of $X$.
For a fixed set $X$ of size $s\le p$, a uniformly random coloring is good for
$X$ with probability
\[
  \frac{p!}{(p-s)!p^s}\ge\frac{p!}{p^p}\ge e^{-p}.
\]
We will derandomize this step after describing the algorithm for one coloring.

Fix $h$ and retain the pairs whose second components are colorful:
\[
  \mathcal P_h:=
  \{(S,T)\in\mathcal P:h(u)\ne h(v)
       \text{ for all distinct }u,v\in T\}.
\]
Enumerate them as $(S_1,T_1),\ldots,(S_r,T_r)$ and write
\[
  C_\ell:=\{h(u):u\in T_\ell\}\subseteq[p]
  \qquad(\ell\in[r]).
\]
For $\ell\in\{0,\ldots,r\}$, $j\in\{0,\ldots,k\}$, and $Q\subseteq[p]$,
let $D_\ell(j,Q)$ be the maximum of
\[
  \sum_{i\in L}w(T_i)
\]
over all $L\subseteq[\ell]$ such that $|L|=j$, the sets $C_i$ for $i\in L$
are pairwise disjoint, and their union is exactly $Q$.  Set
$D_\ell(j,Q)=-\infty$ if no such $L$ exists.  Thus, in plain terms, $\ell$
is the number of pairs processed, $j$ is the number selected, and $Q$ is the
exact set of colors already used; the table entry records the greatest total
weight possible under those conditions.

The initialization is
\[
  D_0(0,\emptyset)=0,
  \qquad
  D_0(j,Q)=-\infty\quad\text{when }(j,Q)\ne(0,\emptyset).
\]
For $\ell\ge1$,
\begin{equation}
\label{eq:bounded-size-dp}
  D_\ell(j,Q)=
  \max\left\{
    D_{\ell-1}(j,Q),
    D_{\ell-1}(j-1,Q\setminus C_\ell)+w(T_\ell)
  \right\},
\end{equation}
where the second term is interpreted as $-\infty$ unless $j\ge1$ and
$C_\ell\subseteq Q$.

We prove the correctness of the recurrence by induction on $\ell$.
For the base case $\ell=0$, the empty selection is the unique selection using
none of the pairs, so it gives $D_0(0,\emptyset)=0$ and every other entry is
$-\infty$, exactly as initialized.  For the induction hypothesis, fix
$\ell\ge1$ and assume that every entry $D_{\ell-1}(j',Q')$ has the meaning
specified above, for all $j'\in\{0,\ldots,k\}$ and $Q'\subseteq[p]$.

For the inductive step, fix $j\in\{0,\ldots,k\}$ and $Q\subseteq[p]$.
Any selection considered by $D_\ell(j,Q)$ either omits or contains the
$\ell$th pair.  If it omits that pair, the induction hypothesis bounds its
value by $D_{\ell-1}(j,Q)$.  If it contains that pair, then $j\ge1$ and
$C_\ell\subseteq Q$; removing the pair leaves $j-1$ pairs whose color sets
have union $Q\setminus C_\ell$, so the induction hypothesis bounds the
remaining value by $D_{\ell-1}(j-1,Q\setminus C_\ell)$.  This gives the two
upper bounds on the right-hand side of \eqref{eq:bounded-size-dp}.
Conversely, a selection attaining the first term remains feasible after the
$\ell$th pair is processed.  When the second term is available, a selection
attaining it uses exactly the colors $Q\setminus C_\ell$, which are disjoint
from $C_\ell$; adding the $\ell$th pair is therefore feasible, gives union
$Q$, and adds $w(T_\ell)$ to its value.  Thus the recurrence computes
$D_\ell(j,Q)$.  Since $j$ and $Q$ were arbitrary, the inductive step holds for
every entry in layer $\ell$, completing the induction.  Recording which term
is chosen in every finite entry permits the selected pairs to be recovered by
traceback.

For the fixed coloring $h$, put
\[
  V_h:=\max_{0\le j\le k}\ \max_{Q\subseteq[p]}D_r(j,Q).
\]
Every selection counted by the table yields feasible pairs: if an element
belonged to two selected second components, its color would belong to both of
their color sets, contradicting their disjointness.  Hence
$V_h\le\OPT_{\mathrm{pack}}(I)$ for every $h$.

For the reverse inequality, fix optimum feasible pairs
$(S_1^*,T_1^*),\ldots,(S_q^*,T_q^*)$ and let
$X:=\bigcup_{i=1}^qT_i^*$.  If $h$ is good for $X$, then every optimum pair
belongs to $\mathcal P_h$, and their color sets are pairwise disjoint.  The
dynamic program can therefore select all these pairs, giving
$V_h\ge\OPT_{\mathrm{pack}}(I)$.  Consequently,
\[
  V_h=\OPT_{\mathrm{pack}}(I)
\]
whenever $h$ is good for the elements used by an optimum solution.

\paragraph*{Derandomization and running time.}
By the standard color-coding derandomization, construct a $p$-perfect hash
family $\mathcal H$, meaning that for every $X\subseteq U$ with $|X|\le p$
some $h\in\mathcal H$ is good for $X$.  Such a family has
$2^{O(p)}\log(2+|U|)$ functions from $U$ to $[p]$ and can be constructed in
$2^{O(p)}N^{O(1)}$ time~\cite{alon1995colorcoding}.  Since an optimum
collection uses at most $kd=p$ elements, some $h\in\mathcal H$ is good for
those elements.  Hence
$\max_{h\in\mathcal H}V_h=\OPT_{\mathrm{pack}}(I)$, and running the dynamic
program for every $h\in\mathcal H$ gives a deterministic exact algorithm.

For a fixed coloring, the table has at most
$(|\mathcal P|+1)(k+1)2^p$ entries.  Representing $Q$ and $C_\ell$ as
$p$-bit vectors makes the subset test and set difference in
\eqref{eq:bounded-size-dp} take $O(p)$ time, so one coloring is processed in
\[
  O\bigl(|\mathcal P|(k+1)2^pp\bigr)
\]
time.  Using $|\mathcal P|\le(2^d-1)|\mathcal S|$, $p=kd$, and
$|\mathcal H|=2^{O(p)}\log(2+|U|)$, the total running time is
$2^{O(kd)}N^{O(1)}$.  Every finite table entry is a sum of at most $kd$ input
element weights, so exact additions and comparisons of the binary rational values
also take polynomial time.

Finally, choose a coloring attaining the maximum and perform the traceback.
This yields feasible pairs of value $\OPT_{\mathrm{pack}}(I)$.  Selecting their
distinct first components produces, by the equivalence above, an optimum
coverage solution and hence an optimum solution to weighted \MinUnCoverage.
\end{proof}